\documentclass[natbib]{svjour3}                     
\smartqed  
\usepackage{graphicx}
\usepackage{paralist}
\usepackage{ctable}
\usepackage{color}
\usepackage{hyperref}
\usepackage{mathptmx}      
\usepackage{aps-bibstyle}  
\usepackage{lscape}

\makeatletter

\newcommand{\Rmnum}[1]{\expandafter\@slowromancap\romannumeral #1@}
\makeatother
 \journalname{Space Science Reviews}
\newcommand{\adv}{Adv. Space Res.}

\newcommand{\aap}{Astron. Astrophys.}
\newcommand{\aaps}{Astron. Astrophys. Suppl.}

\newcommand{\an}{Astron. Nachr.}

\newcommand{\aj}{Astron. J.}
\newcommand{\apj}{Astrophys. J.}

\newcommand{\apjs}{Astrophys. J. Supplement}

\newcommand{\grl}{Geophys. Res. Lett.}

\newcommand{\jgr}{J. Geophys. Res.}

\newcommand{\pasp}{Publ. Astron. Soc. Pacific}
\newcommand{\pasj}{Pub. Astron. Soc. Japan}

\newcommand{\procspie}{Proc. SPIE}
\newcommand{\solphys}{Sol. Phys.}

\newcommand{\ssr}{Space Sci. Rev.}
\newcommand{\planss}{Planet. Space Sci.}

\begin{document}


\title{Large-scale Bright Fronts in the Solar Corona: A Review of ``EIT waves"}
\titlerunning{A Review of ``EIT waves"}      

\author{Peter T. Gallagher \and David M. Long}
\institute{P.T. Gallagher \and D.M. Long \at
	         Astrophysics Research Group, School of Physics, Trinity College Dublin, Dublin 2, Ireland.\\
	         \email{peter.gallagher@tcd.ie}}

\date{Received: date / Accepted: date}

\maketitle


\begin{abstract}

``EIT waves" are large-scale coronal bright fronts (CBFs) that were first observed in 195~\AA\ images obtained using the Extreme-ultraviolet Imaging Telescope (EIT) onboard the \emph{Solar and Heliospheric Observatory (SOHO)}. Commonly called ``EIT waves", CBFs typically appear as diffuse fronts that propagate pseudo-radially across the solar disk at velocities of 100--700~km~s$^{-1}$ with front widths of 50--100~Mm. As their speed is greater than the quiet coronal sound speed ($c_s\leq$200~km~s$^{-1}$) and comparable to the local Alfv\'{e}n speed ($v_A\leq$1000~km~s$^{-1}$), they were initially interpreted as fast-mode magnetoacoustic waves ($v_{f}=(c_s^2 + v_A^2)^{1/2}$). Their propagation is now known to be modified by regions where the magnetosonic sound speed varies, such as active regions and coronal holes, but there is also evidence for stationary CBFs at coronal hole boundaries. The latter has led to the suggestion that they may be a manifestation of a processes such as Joule heating or magnetic reconnection, rather than a wave-related phenomena. While the general morphological and kinematic properties of CBFs and their association with coronal mass ejections have now been well described, there are many questions regarding their excitation and propagation. In particular, the theoretical interpretation of these enigmatic events as magnetohydrodynamic waves or due to changes in magnetic topology remains the topic of much debate.


\end{abstract}


\section{Introduction}
\label{sect:intro}

The solar corona is a hot ($T_e=$1--2~MK), tenuous ($n_e\sim$10$^9$~cm$^{-3}$) plasma threaded by magnetic fields ($B\sim$10--1000~G), which emits strongly at extreme-ultraviolet (EUV) and X-ray wavelengths. With recent advances in ground and space-based instrumentation, it has become clear that the corona is ever-changing, evolving on time scales ranging from milliseconds (e.g., radio spikes) to months (e.g., coronal streamers). Among the most spectacular manifestations of solar activity are solar flares and coronal mass ejections (CMEs) which result from rapid changes in the coronal magnetic field \citep[e.g.,][]{priestNforbes2000}. This rapid, large-scale re-organisation and relaxation of non-potential magnetic fields can produce 10$^{25}$~J in a matter of minutes, and are frequently accompanied by a multitude of associated transient phenomena, including shocks, metric Type II and Type III radio bursts, solar energetic particle events, and globally-propagating wave-like features, such as Moreton waves in the chromosphere and ``EIT waves" in the corona.

Globally-propagating disturbances were first observed on the Sun in the chromospheric H$\alpha$ line during the early 1960s \citep{moreton1960}. Since then, similar large-scale propagating disturbances have been observed in images obtained at extreme ultraviolet (EUV), soft X-ray, He~{\sc i}~(10830~\AA), and radio wavelengths. In the EUV, large-scale disturbances are commonly called ``EIT waves" after their first observation  \citep{moses1997,thompson1998} using the Extreme-ultraviolet Imaging Telescope \citep[EIT;][]{delaboudiniere1995} instrument onboard the \emph{SOlar and Heliospheric Observatory} \citep[SOHO;][]{domingo1995}. These large-scale coronal bright fronts (CBFs) generally appear as broad, diffuse features that form a single arc-shaped front when unimpeded in the quiet solar atmosphere, which maintain their coherence across large length-scales ($\leq$1~R$_\odot$). Their measured speed varies over a wide range, from 100--700~km~s$^{-1}$, which is somewhat below the coronal Alfv{\'e}n speed and lead a number of authors to interpret them as magnetoacoustic waves \citep[e.g.,][]{thompson1998,wang2000}. It is now commonly believed that these low measured speeds may have resulted from systematic under-sampling of the CBF velocity profile \citep[e.g.,][]{long2008}. The wave-like behavior of CBFs is further evident in areas of the solar atmosphere where coronal plasma conditions vary (e.g., in active regions and coronal holes). Here, CBFs have been found to experience reflection and refraction \citep{gopalswamy2009}, although they have been observed to remain stationary at the boundary of coronal holes \citep{delannee1999}. From a theoretical perspective, this has led to some difficulty in interpreting their physical origin \citep{chen2005,attrill2006,crookerNwebb2006}. Observationally though, the morphological and kinematic properties of CBFs, together with their relationship with other transient phenomena, such as flares, Type II radio bursts and coronal mass ejections (CMEs), have now been systematically studied for over a decade; these are collated in Table~\ref{table:observations}.

\ctable[
caption = {Observational properties of ``EIT waves" or CBFs. Note that these properties have been taken from multiple sources which interpret CBFs differently.  They are therefore contradictory in some instances (e.g., Gaussian pulse versus wave train).},
label = table:observations,
center,
width=\textwidth,
pos = h!
]{ll}{
\tnote[1]{\citet{willsdaveyNthompson1999}, \citet{thompson1998}, \citet{zhukov2004}, \citet{long2008}}
\tnote[2]{Temperature of coronal plasma imaged by the passbands noted above.}
\tnote[3]{Height obtained using \emph{STEREO} quadrature observations by \citet{patsourakos2009,kienreich2009}.}
\tnote[4]{Noted from first observations of an CBF by \citet{thompson1999}.}
\tnote[5]{Measured by \citet{willsdavey2006} using cross-sections of an CBF front.}
\tnote[6]{CBFs observed to avoid both active regions and coronal holes\citet{thompson1999}.}
\tnote[7]{Quantified and measured by \citet{gopalswamy2009}}
\tnote[8]{Observed in simulations \citep{ofmanNthompson2002,wang2000}.}
\tnote[9]{\citet{willsdavey2007,ballai2005}}
\tnote[10]{\citet{warmuth2004b}}
\tnote[11]{\citet{thompson1999}}
\tnote[12]{\citet{willsdavey2006,veronig2010}}
\tnote[13]{\citet{thompson2009}}
\tnote[14]{\citet{ballai2005}}
\tnote[15]{Based on mean wave velocities and periods.}
\tnote[16]{Comparable to a nanoflare \citep{ballai2005}.}
\tnote[17]{\citet{warmuth2004b,long2008,veronig2008}}
\tnote[18]{\citet{thompson2009,willsdavey2007,wang2000,chen2002}.}
\tnote[19]{\citet{klassen2000,warmuth2001,warmuth2004a}.}
}{
\FL
Properties & Observations \ML
Passbands & 171~\AA\tmark[1], 195~\AA\tmark[1], 284~\AA\tmark[1], 304~\AA\tmark[1] \NN
Temperature range &  1--2~MK\tmark[2] \NN
Height range & $\sim$70--90~Mm layer above photosphere\tmark[3] \NN
Wavefront shape & Anisotropic and non-homogeneous\tmark[4] \NN
Pulse shape & Gaussian-like\tmark[5] \NN
Region of atmosphere & Quiet Sun\tmark[6] \NN
Reflection & At active region/coronal hole boundaries\tmark[7] \NN
Refraction & At active region/coronal hole boundaries\tmark[8] \NN
Dispersion & Minimal\tmark[9]/small\tmark[10] \NN
Amplitude & $\sim$14--23~\%\tmark[11] \NN
FWHM & $\sim$25--100~Mm\tmark[12] \NN
Velocity & $\sim$20--600~km/s\tmark[13] \NN
Period & $\sim$400~s\tmark[14] \NN
Wavelength & 100--300~Mm\tmark[15] \NN
Energy & $\sim$10$^{18}$~J\tmark[16] \NN
Acceleration & Deceleration observed by multiple authors\tmark[17] \NN
Alfv\'{e}n Mach number & $<$1\tmark[18] \NN
 & $>$1\tmark[19] \LL
}

The complex behavior of CBFs has given rise to two conflicting sets of theories, one proposing that they are true waves, the other proposing that they are pseudo-waves. The wave theory treats CBFs as fast-mode magnetohydrodynamic (MHD) waves or shock waves excited by a solar flare or erupting coronal mass ejection \citep[e.g.,][]{uchida1968,warmuth2004b}, while \citet{willsdavey2007} interpreted them as soliton waves. More recently, numerical simulations by \citet{wang2009} have shown that both slow- and fast-mode waves can be produced by an erupting flux rope, which have characteristics similar to CBF. Alternatively, pseudo-wave theories treat CBFs as a result of the global restructuring of the solar magnetic field during a CME eruption \citep[e.g.,][]{chen2002,chen2005,attrill2007,delannee2008}. A number of these theories postulate that the bright front of an CBF results from magnetic reconnection along the expanding flanks of a CME. To date, neither the wave nor the pseudo-wave theories have succeeded in explained the various observational properties of CBFs. Unfortunately, the link between CBFs and CMEs in the low corona can only be investigated observationally using simultaneous on-disk EUV and near-Sun coronagraph images ($\leq$2~$R_\odot$), which has not been possible since the loss of the inner coronagraph (C1) of the Large Angle and Spectrometric Coronagraph \citep[LASCO;][]{brueckner1995} on \emph{SOHO}. 

The launch of the \emph{Solar TErrestrial RElations Observatory (STEREO)} spacecraft have enabled a more complete analysis of CBFs, due to the higher cadence of their EUV imagers and the availability of two distinct viewpoints \citep[e.g.,][]{kienreich2009}. The ability to reconstruct the three-dimensional (3D) structure of CBFs using the two perspective views of \emph{STEREO} has given us a new understanding of their morphology, temporal evolution, and relationship with CMEs \citep{patsourakos2009}. The launch of the ESA \emph{Proba-2} mission with the Sun Watcher using Active pixel system detector and image Processing \citep[SWAP;][]{berghmans2006} in November 2009, and NASA's \emph{Solar Dynamic Observatory (SDO)} in February 2010 now enable us to study these phenomena in greater detail than ever before. \emph{SDO} in particular is providing a revolutionary new insight into the multi-thermal signature of CBFs with unprecedented temporal and spatial resolution. The excellent sensitivity and resolution of these imagers may in fact enable us to use CBFs to probe plasma conditions in the solar corona through the use of coronal seismology \citep[e.g.,][]{ofmanNthompson2002,ballai2005}.

In this review, we summarize the evolution in our understanding of CBFs, from their discovery and naming as ``EIT waves" in 1997, to recent quadrature observations using \emph{STEREO} in 2010. In Section~\ref{sect:morphology} we discuss the observational properties of these disturbances, focusing on their morphology as inferred from a number of EUV imagers, while their kinematics are described in Section~\ref{sect:kinematics}. In Sections~\ref{sect:chrom_sig} and \ref{sect:corona_sig}, we review related wave-like phenomena in the chromosphere and corona, while Section~\ref{sect:radio} includes a discussion of radio emission associated with CBFs. The interaction of CBFs with other coronal magnetic features and their association with flares and CMEs are then discussed in Sections~\ref{sect:interaction} and \ref{sect:flare_cme}. The theories attempting to explain CBFs are then presented in Section~\ref{sect:modelling}. Finally, Section~\ref{sect:conclusions} includes a discussion of prospects for improving our understanding of CBFs and raise questions which any theory purporting to explain these phenomena must address. The reader may be interested in complementary reviews by \citet{willsdaveyNattrill2010} and \citet{warmuth2010,warmuth2007}. 

\section{Coronal Bright Front Morphology}
\label{sect:morphology}

\begin{figure*}[!t]
\includegraphics[width=1\textwidth]{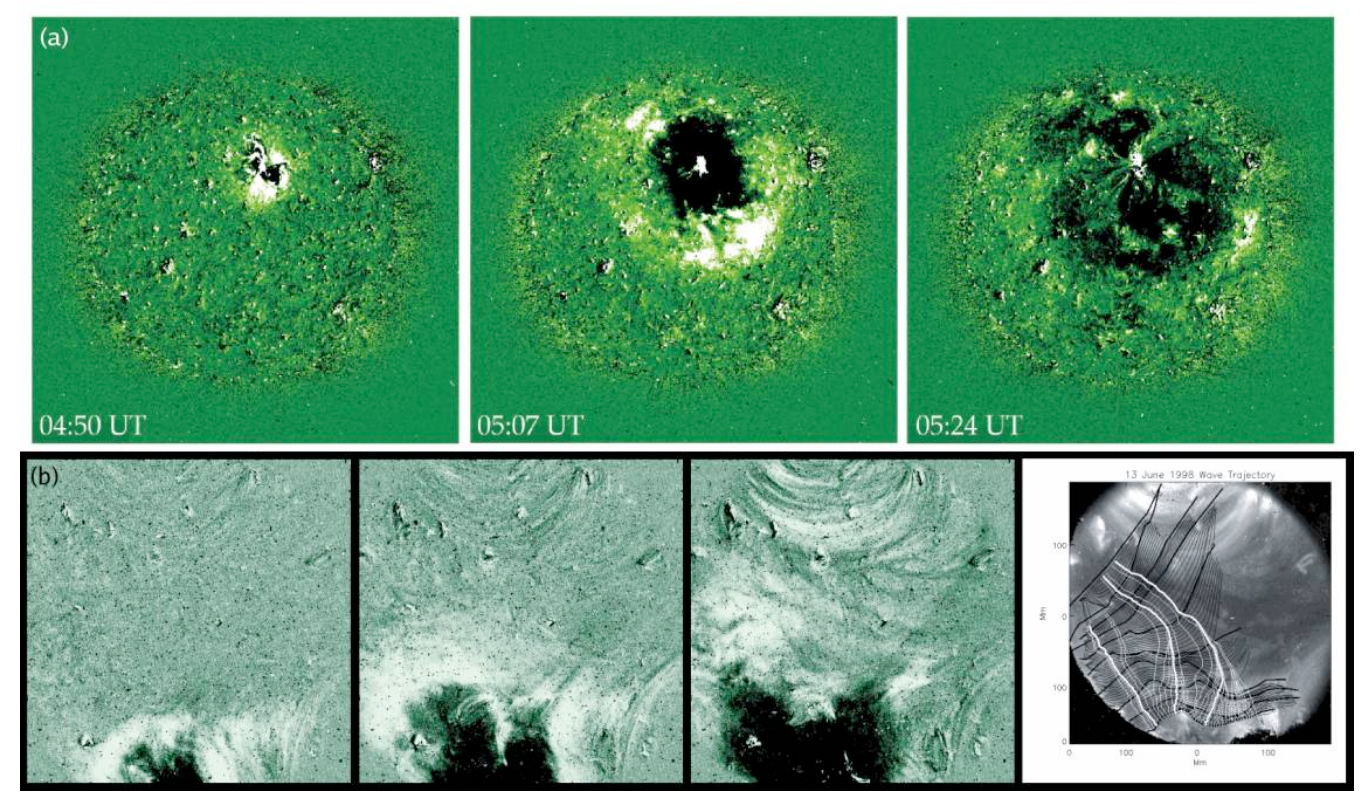}
\caption{Two examples of CBFs, commonly called ``EIT waves", as observed by different EUV instruments \citep{willsdavey2007}. (a) Running difference images of a CBF seen by \emph{SOHO}/EIT on 12~May~1997, and studied in detail by \citet{thompson1998}. (b) Base difference images and measured fronts from \citet{willsdavey2006} of an event observed by \emph{TRACE} on 13~June~1998.}
\label{fig:willsdavey2007}       
\end{figure*}

\citet{thompson1998} were the first to carry out a detailed study of CBFs, which were named ``EIT waves" for the instrument in which they were first observed.\footnote{As these phenomena are observed across multiple coronal passbands and have not been conclusively shown to be waves, ``coronal bright front (CBF)" is a more appropriate name. The fact that they have also been observed to remain stationary for many tens of minutes, also argues against the use of the words ``propagating" or ``wave" in their title.} They reported several CME signatures in EIT 195~\AA\ images, including dimming regions close to the eruption, post-eruption arcade formation, and a {\it bright wavefront} propagating quasi-radially from the source region (see the top row of Figure~\ref{fig:willsdavey2007}). The propagating bright wavefront became known as an ``EIT wave". Each of the phenomena discussed by \citet{thompson1998} appeared to be associated with the same eruption, with the onset time of the various features corresponding with the estimated onset time of the CME inferred using images from \emph{SOHO}/LASCO. In a follow-on paper, \citet{thompson1999} suggested that CBFs could be strong candidates for the coronal manifestation of Moreton waves. They found that the relatively weak amplitudes of these waves ($\sim$14-25\% above background) indicated that the coronal wave may not always be shock-like in nature. Furthermore, the wave fronts were reported to be diffuse and not to display distinct shock boundaries. On the basis of speed and propagation characteristics, they concluded that there was strong circumstantial evidence for the association of an CBF with the Moreton wave phenomenon.

\begin{figure*}[!t]
\begin{center}
\includegraphics[width=0.9\textwidth]{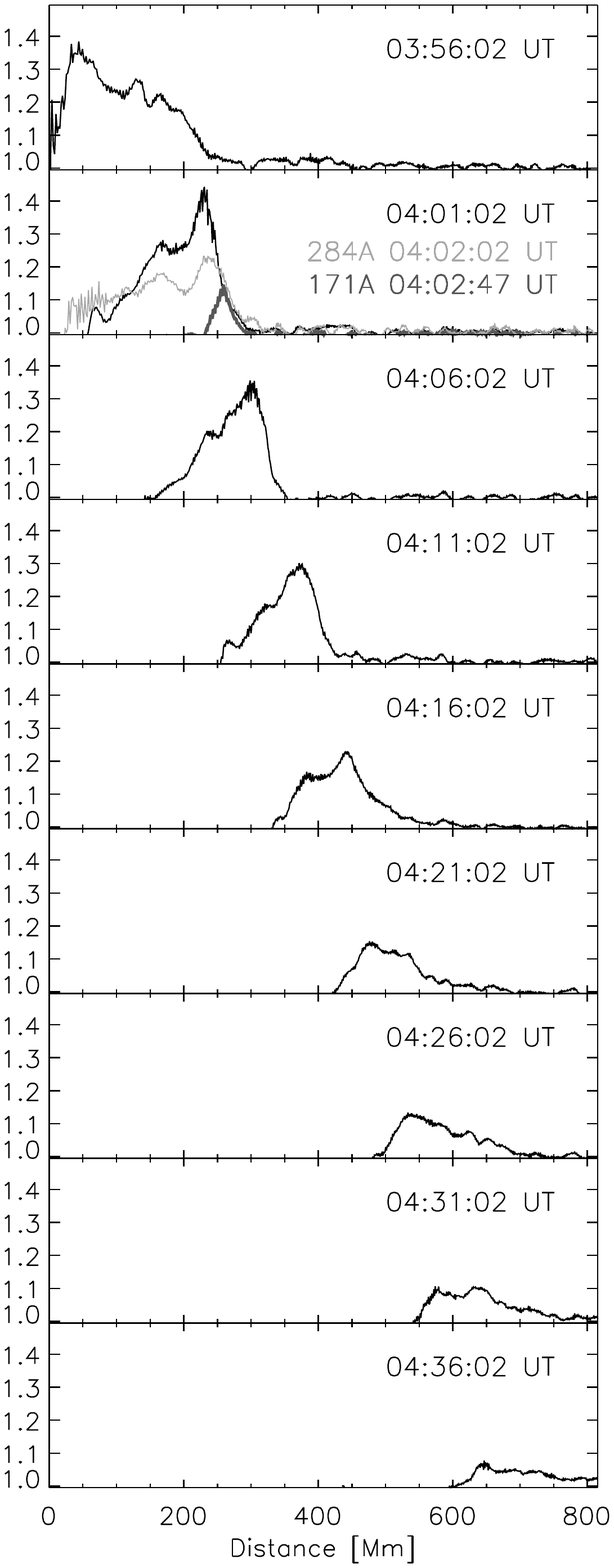}
\caption{\emph{STEREO}/EUVI 195~\AA\ CBF pulse intensity profiles. The profiles were created by summing over 60 degree wide sectors in running ratio images. The image cadence was 10 minutes.  Following launch, the pulse initially steepens as the amplitude increases. Thereafter, it can be seen to decrease in amplitude and broaden as it propagates away from the source, although the integrated area of the pulse profile remains constant with time. See \citet{veronig2010} for further details.}
 \label{fig:veronig2010}
\end{center}
\end{figure*}

\citet{willsdaveyNthompson1999} identified a similar propagating disturbance using \emph{Transition Region and Coronal Explorer} \citep[TRACE;][]{handy1999} 171 and 195~\AA\ passbands images. While the 171~\AA\ passband exhibited strong displacement of individual magnetic structures, the 195~\AA\ images revealed a strong wave front and associated dimming but resolved much less structural motion. There was also some evidence for heating in the wave front, while a comparison of the 171 and 195~\AA\ images enabled the authors to constrain the temperature of the plasma through which the wave propagates to 1--1.4~MK. Variation in the trajectories and velocities of points along the front led \citet{willsdaveyNthompson1999} to suggest that the disturbance was Alfv\'{e}nic in nature and contained a compressive component. By examining intensity cross-sections of the front, they showed that the density perturbations exhibit a roughly Gaussian wave structure, suggesting a single propagating compression front. The wave fronts were also found to propagate nonuniformly, unlike the near-circular fronts often seen with \emph{SOHO}/EIT. The roughly Gaussian pulse shapes of CBFs together with their apparent pulse coherence and sub-Alfv\'{e}nic velocities lead \citet{willsdavey2007} to postulate that they were more consistent with solitons that fast-mode MHD waves, although it was not until the recent work of \citet{veronig2010} that the pulse shape could be studied with any certainty. Using 195~\AA\ passband images from the Extreme UltraViolet Imager \citep[EUVI;][]{wuelser2004} on \emph{STEREO}, \citet{veronig2010} found convincing evidence for decreasing amplitude over $\sim$40-minutes, while the outer edge of the wave front is steepest at the time of maximum amplitude (see Figure~\ref{fig:veronig2010}). These observations are particularly compelling evidence for the wave interpretation of CBFs.

The evidence for CBFs being actual waves was questioned by \citet{delannee1999} and \citet{delannee2000}. \citet{delannee1999} found that a bright front can lie at the same location for several hours, which was taken as strong evidence for the implausibility of the wave interpretation. A number of dimmings associated with neighboring magnetic structures, such as transequatorial loops and the source active region, were also reported. It was proposed that these features were strongly related to magnetic field topology. This analysis led \citet{delannee2000} to conclude that that the CBF phenomenon is more closely related to the magnetic field evolution involved in CMEs than to wave propagation driven by solar flares. Using an automated algorithm to detect and study CBFs, \citet{podladchikova2005} presented evidence for angular rotation of CBFs with propagation. This result was confirmed by \citet{attrill2007}, who investigated the properties of two classical CBFs, reporting deep core dimmings near the flare site and also widespread diffuse dimming accompanying the expansion of the CBF. They also report a new property of these CBFs, namely that they display dual brightenings: persistent ones at the outermost edge of the core dimming regions and simultaneously diffuse brightenings constituting the leading edge of the coronal wave, surrounding the expanding diffuse dimmings. In keeping with the suggestions of \citet{delannee2000}, they postulated that such behavior is consistent with a diffuse CBF being the magnetic footprint of a CME. Interestingly, \citet{thompson2000} suggested the MHD interpretation and the opening magnetic field line approach may both be able to explain many features of the 24~September~1997 observations.

\citet{biesecker2002} investigated the wave and pseudo-wave interpretations of CBFs using a catalogue of 173 events published by \citet{thompson2009}. Since CBFs are transient, coronal phenomena, they searched for correlations with other transient, coronal phenomena, such as X-ray flares, CMEs, and Type II bursts. A clear correlation between CBFs and CMEs was found, while a their correlation with solar flares was significantly weaker. Furthermore, CBFs  were not frequently found to be accompanied by radio bursts, casting doubt on the shock interpretation of these events. According to \citet{biesecker2002}, the majority of the CBF reported in the Thompson \& Myers catalogue consist of diffuse brightenings of relatively low amplitude. However, a very small fraction of the CBF (7\%) had sharp, bright components associated with them. In two of the cases where the CBF had a sharp, bright feature associated with them, other studies \citep[cf.][]{warmuth2004a} have shown that there was a correlated Moreton wave. Thus, the sharp, bright feature, only occasionally seen in CBFs, may be a signature of a Moreton wave. 

\begin{landscape}
\begin{figure*}
\begin{center}
\includegraphics[height=0.8\textwidth]{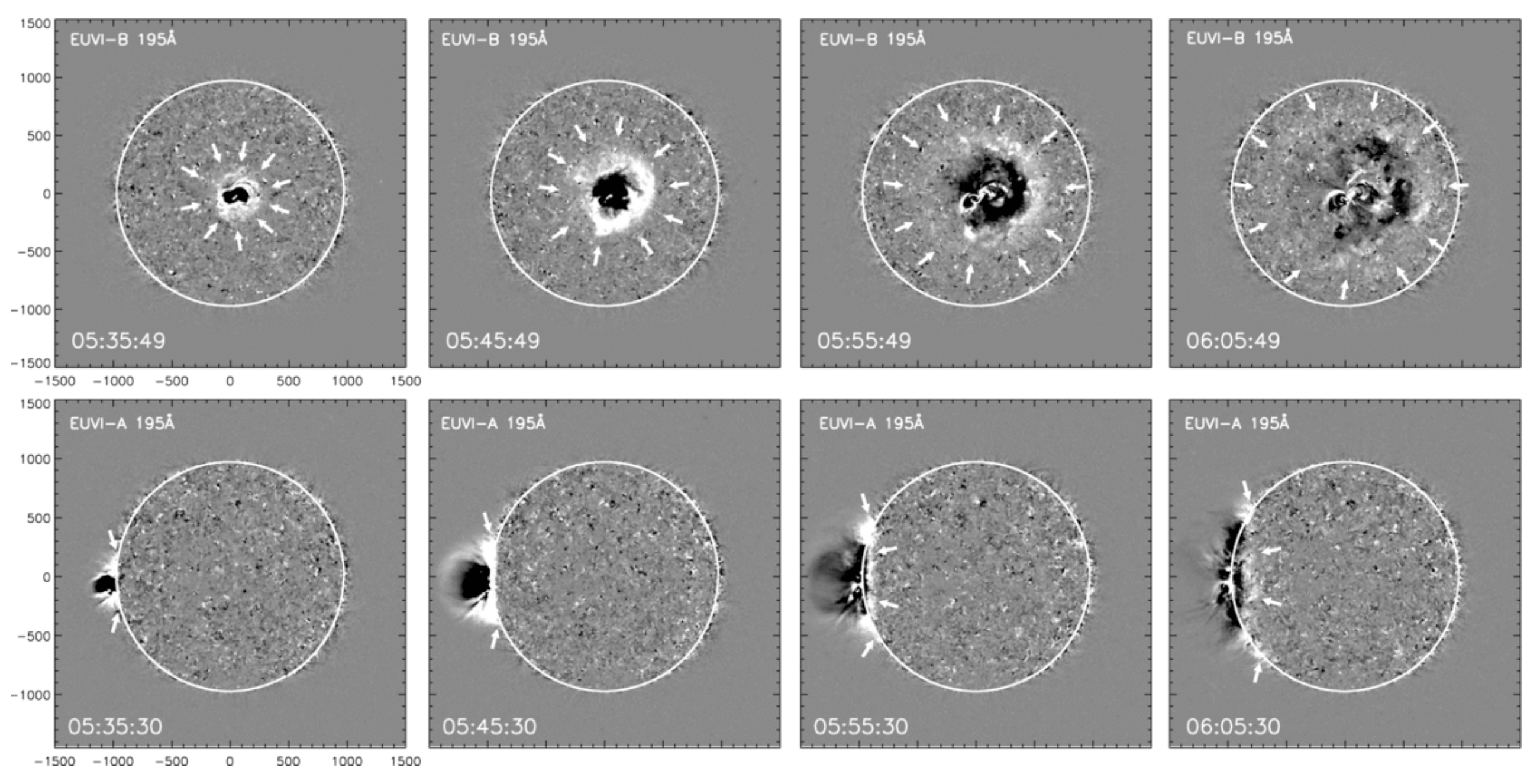}
\caption{Sequence of median-filtered running difference images recorded in the EUVI 195~\AA\ channel with a cadence of 10 minutes from \citet{kienreich2009}. The coronal wave (outlined by arrows) is observed on-disk in \emph{STEREO}-B (top row) and on the limb in \emph{STEREO}-A (second row).}
\label{fig:kienreich_09}       
\end{center}
\end{figure*}
\end{landscape}

\begin{figure*}[!t]
\begin{center}
\includegraphics[width=1\textwidth,trim = 0mm 50mm 0mm 50mm,clip=]{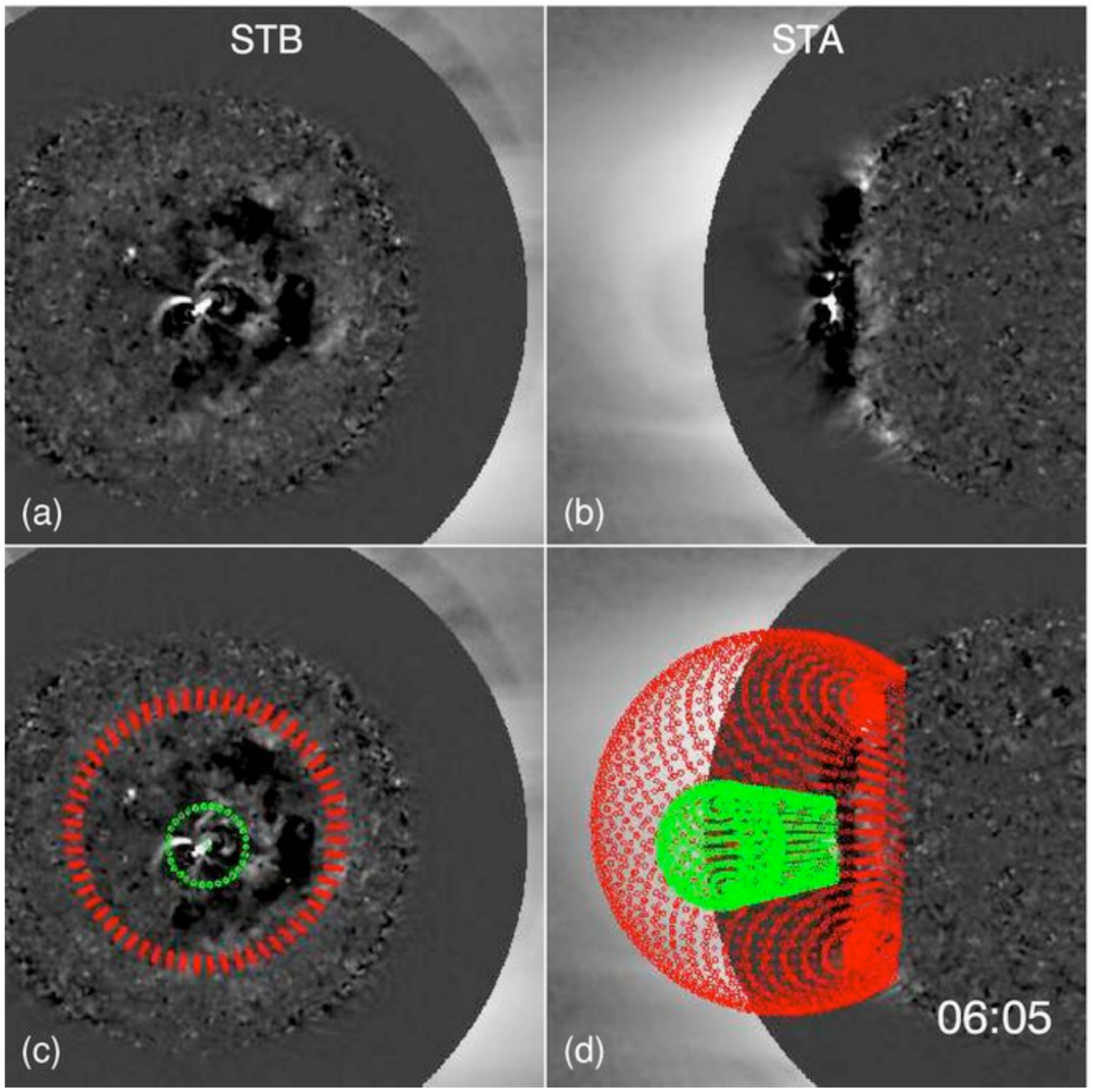}
\caption{Forward modeling of the CME and wave 3D shape in both \emph{STEREO} spacecrafts \citep{patsourakosNvourlidas2009}. Panels (a) and (b) contain composite EUVI 195~\AA\ running difference and COR1 images from \emph{STEREO}-B and \emph{STEREO}-A. Panel (d) contains the best-fit CME (green wireframe) and wave (red wireframe) model determined for \emph{STEREO}-A. Panel (c) has the disk projections of these models in \emph{STEREO}-B. This is the same event as shown in Figure~\ref{fig:kienreich_09} \citep{kienreich2009}.}
\label{fig:patsourakos_09}       
\end{center}
\end{figure*}

A new insight into the 3D structure and morphology of CBFs has recently been gained using the different perspectives offered by the \emph{STEREO} spacecraft.  \citet{kienreich2009} and \citet{patsourakosNvourlidas2009} studied the basic 3D structure of an expanding CBF, finding it to have a thin dome-like structure.  \citet{kienreich2009} used lateral observations of the wave from  \emph{STEREO}-A to study the 3D nature of CBFs, with the wave front visible to a considerable height above the solar surface (see Figure~\ref{fig:kienreich_09}). Since \emph{STEREO}-A has an edge-on view of the wave, the distance measurements are free from projection effects. They found that the resulting disturbance positions lie 50--100~Mm behind the kinematical curve derived from the on-disk \emph{STEREO}-B observations, indicating that the wave propagates significantly above the solar surface. This indicates that the EUV wave emission observed on-disk originates from high above the solar surface of 80--100~Mm. This is similar to the height of 90$\pm$7~Mm derived for a global EUV wave observed with \emph{STEREO} at a spacecraft separation of 45 degrees using triangulation techniques \citep{patsourakosNvourlidas2009}. Such heights are comparable to the coronal scale-height of 50--100~Mm for quiet Sun temperatures of 1--2~MK, and typical of what is expected for the propagation of a fast-mode MHD wave in the quiet solar corona. From a consideration of the associated CMEs flank positions and expansion, they interpreted the CBF as a fast-mode wave driven forward of the CME legs. \citet{patsourakos2009} investigated the CME-wave relation using a simple geometric model (see Figure~\ref{fig:patsourakos_09}). The projection of the front from their geometric model agrees well with the observations, while the CME projection is smaller and confined around the active region core dimming. The forward modelling suggests that the wave and the CME are not concentric with each other or with the active region center. More recently, \citet{veronig2010}  presented strong evidence that EUV coronal waves have a dome-shaped geometry, arguing that the observed structure is a wave dome and not a CME. This was primarily motivated by the spherical form and sharpness of the domeÕs outer edge, and by the fact that the the low-coronal wave signatures above the limb connect near perfectly to the on-disk signatures of the wave. These spectacular 3D observations from \emph{STEREO} have not only provided us with a new view of CBF  morphology and its temporal variation, but has enabled us to study their motion, both parallel and perpendicular to the solar surface. The detailed kinematics of CBFs as revealed from observations is discussed in more detail in the next section.

\section{Coronal Bright Front Kinematics}
\label{sect:kinematics}

Early observations reported CBFs with constant velocities of several hundred kilometres per second \citep{thompson1998}. Initially, the disturbances were found to be in excess of the estimated coronal sound speed and somewhat below the Alfv\'{e}n speed, with \citet{thompson2009} reporting a large range of velocities, from 50--700~km~s$^{-1}$, although they did state that a typical velocity is more in the range 200--400~km~s$^{-1}$. This is comparable to the typical range of velocities reported for CBFs by \citet[][170--350~km~s$^{-1}$]{klassen2000}. Using the higher image cadence of \emph{TRACE} ($\sim$1~minute), albeit at the expense of field-of-view, \citet{willsdaveyNthompson1999} determined the velocity of a number of portions of a CBF-like disturbance.  Most of the velocities were between 200--800~km~s$^{-1}$. Such a range of velocities led them to speculate that the wave had a strong Alfv\'{e}nic component and was reacting to variations in the medium through which it was traveling. The upper limit to their observed velocities led them to conclude that a fast-mode magnetoacoustic interpretation was most consistent with the data. 

\begin{figure*}[!t]
\begin{center}
\includegraphics[width=0.9\textwidth]{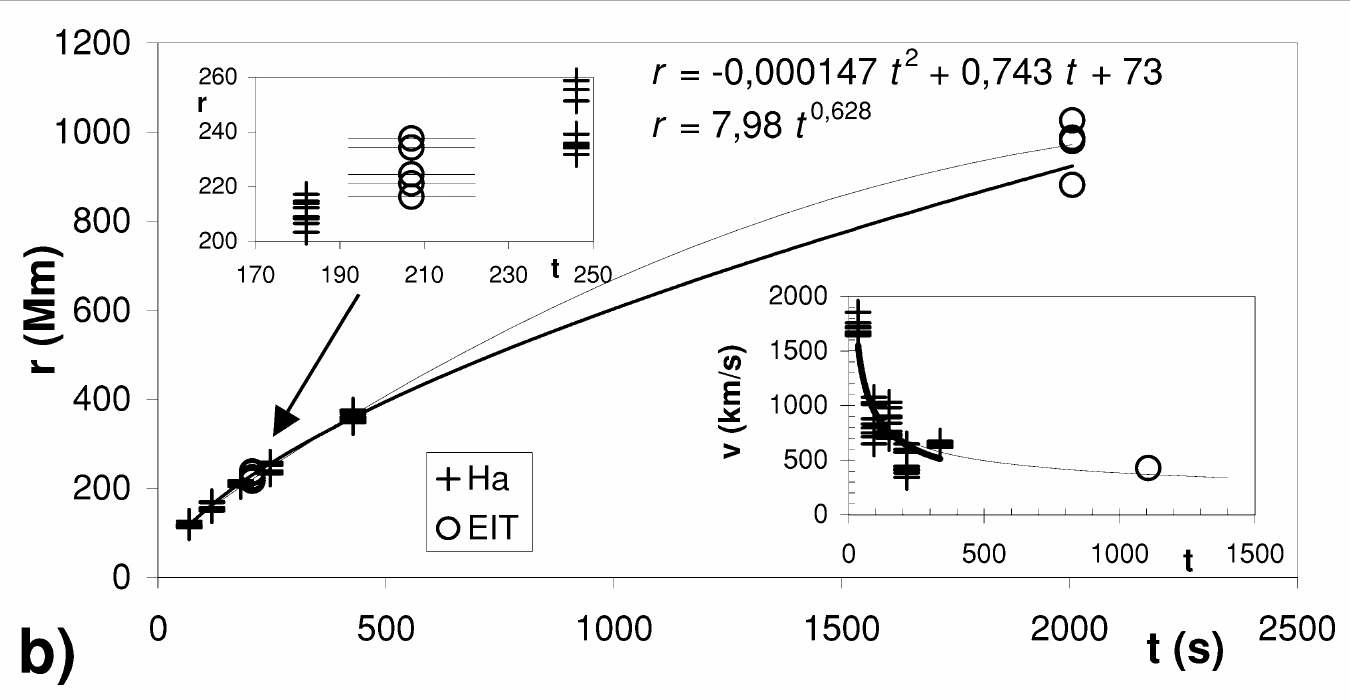}
\caption{Propagation of a CBF and Moreton wave studied by \citet{warmuth2001}. In the inset above the curves, an zoom-in shows the close association of the H$\alpha$ front and CBF observed using EIT. Power-law (thick curve) and second-degree polynomial (thin curve) fits are shown. The inset at bottom right shows the velocity as a function of time from H$\alpha$ (crosses) and EIT (circles).}
\label{fig:warmuth2001}
\end{center}
\end{figure*}

The constant velocity model of CBF  motion has been investigated using recent 3D reconstructions using images from \emph{STEREO}/EUVI \citep{kienreich2009,ma2009}. \citet{kienreich2009} found that the CBF propagates globally over the visible solar hemisphere with a constant velocity of 263$\pm$16~km~s$^{-1}$. Comparison of the wave kinematics with the early phase of the associated CME indicates that the wave is initiated by the CME lateral expansion, and then propagates freely with a velocity close to the fast magnetosonic speed in the quiet solar corona. For the same event, \citet{patsourakosNvourlidas2009} found that the wave kinematics could be best fit by a non-zero acceleration of -25~m~s$^{-2}$. Further evidence for the constant velocity interpretation of CBFs has been given in a recent paper by \citet{veronig2010}. Using limb observations of a dome-shaped CBF, they found a lateral expansion velocity of $\sim$280~km~s$^{-1}$. In addition, they showed that the wave dome expanded in an upward direction relative to the solar surface with a much greater velocity ($\sim$650~km~s$^{-1}$). 

A number of authors have found that the constant velocity assumption is not in fact consistent with the observed kinematics of CBFs \citep{warmuth2001,vrsnak2002,veronig2008,long2008}. Using two events observed in EIT 195~\AA\, supplemented by H$\alpha$ images, \citet{warmuth2001} reported a non-linear velocity profile, implying a non-zero acceleration for the disturbance for at least the first $\le$2000~s of the observations. Figure~\ref{fig:warmuth2001} shows the distance as a function of time for a well observed CBF in both EIT 195~\AA\ and H$\alpha$ \citep{warmuth2001}. Also shown are quadratic (thin) and power-law (thick) fits to the distance-time data. The resulting velocity profiles are given in the inset at bottom right of the figure. It is clear from the figure that the constant acceleration model is a good fit to both the distance- and velocity-time measurements. This led \citet{warmuth2001} to conclude that CBFs and Moreton waves are signatures of the same propagating wavefront. In a follow-on paper, \citet{warmuth2004a} studied twelve flare-associated wave events using data-sets from different passbands (EUV, He~{\sc i}~10830~\AA, SXR and radioheliographic data). Here they found that the wave signatures in the various spectral bands lie on closely associated kinematical curves, implying that they are signatures of the same physical disturbance. In all the events studied, and at all wavelengths, the waves were observed to be decelerating, which was used to explain the apparent velocity discrepancy between Moreton waves and CBFs. The conclusion drawn using the available data-sets is that this behaviour may be explained by a freely-propagating fast-mode MHD shock formed from a large amplitude single wave. Similarly, \citet{vrsnak2002} found evidence for deceleration of the order of 100--1000~m~s$^{-2}$ in chomospheric images centred on He~{\sc i}~ (10830~\AA) and H$\alpha$~(6563~\AA). Similar decelerations were found for the associated CBF observed in \emph{SOHO}/EIT (195~\AA) passband images.

\begin{figure}[t!]
\begin{center}
\includegraphics[keepaspectratio, width=1\textwidth,trim = 0mm 0mm 0mm 0mm,clip=]{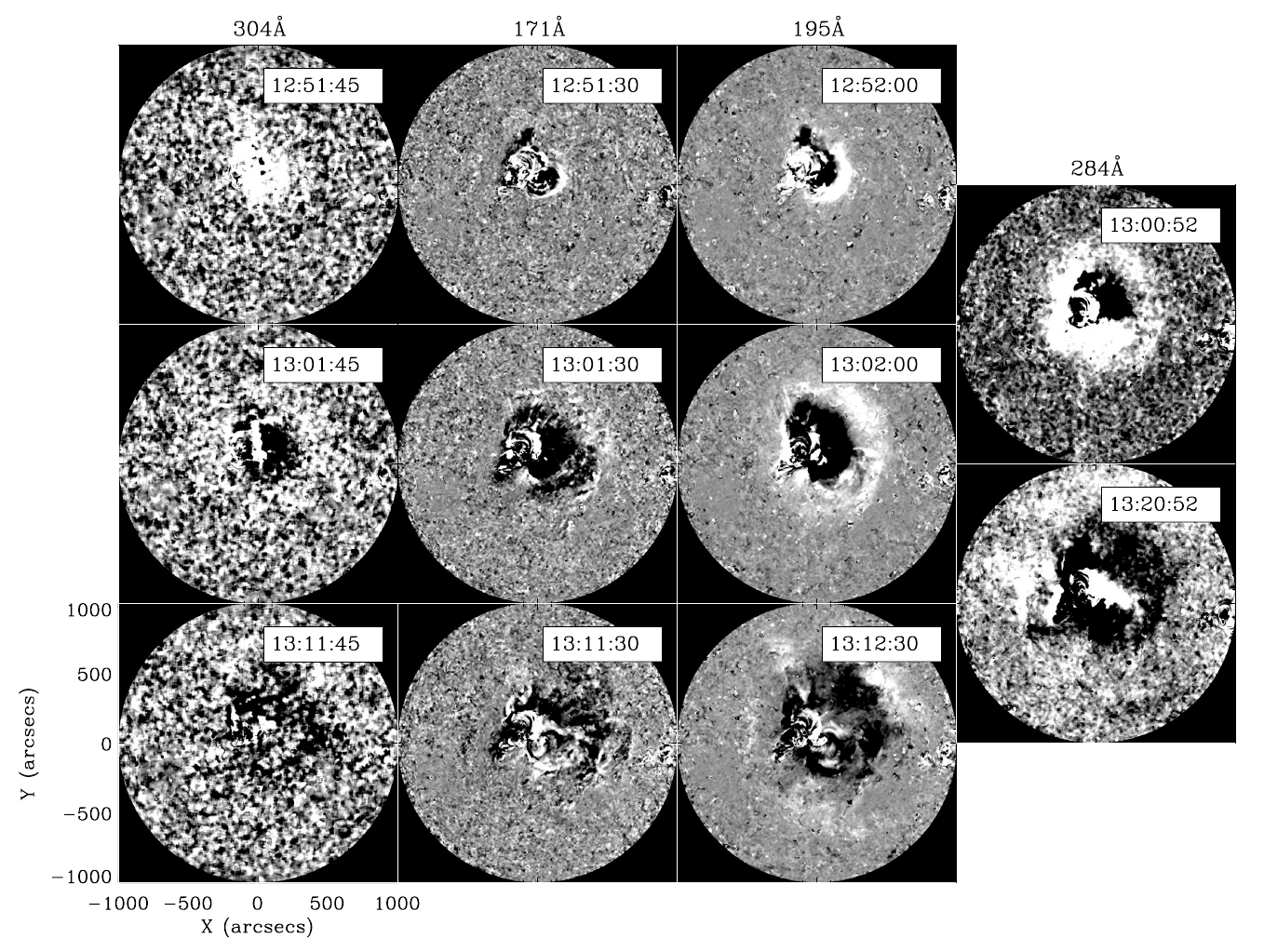}
\caption{Evolution of a CBF over approximately 20 minutes in all four \emph{STEREO/}EUVI wavebands. Time increases from top to bottom, while the wavelengths are arranged from left to right as 304, 171, 195, and 284~\AA. See \citet{long2008} for a more detailed discussion.}
\label{fig:longetal}
\end{center}
\end{figure}

\begin{figure*}[!t]
\begin{center}
\includegraphics[width=0.9\textwidth]{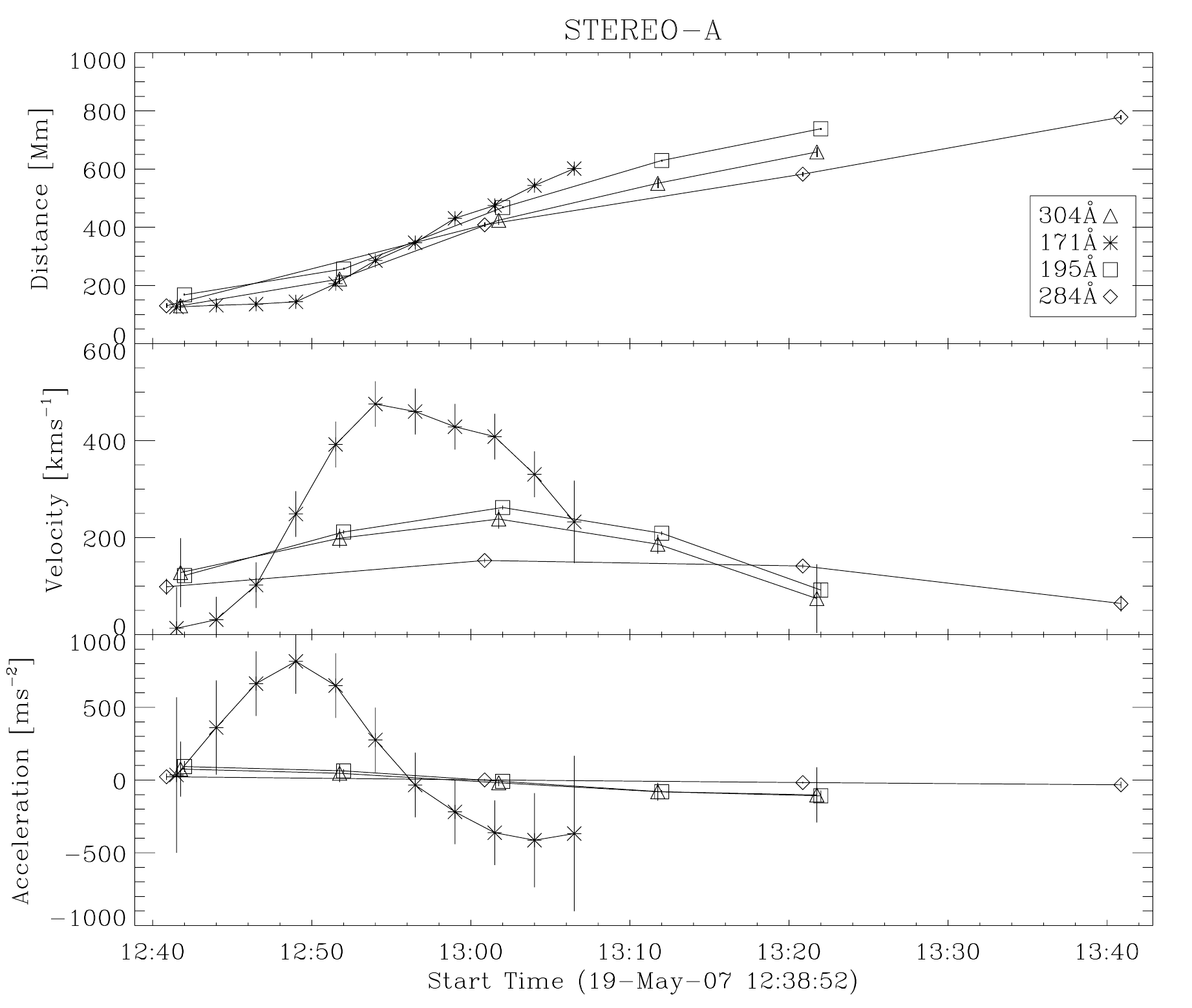}
\caption{Distance-time (top), velocity-time (middle), and acceleration-time (bottom) plots for a CBF observed by \emph{STEREO}/EUVI. Distances are measured from the flare kernel to the leading edge of the running difference brightening along the great circle longitude line to solar west \citep[][]{long2008}.}
 \label{fig:long2008}
\end{center}
\end{figure*}

\citet{long2008} investigated the non-constant velocity of CBFs using 2.5--10 minute cadence images of the 19~May~2007 event from \emph{STEREO}/EUVI (see \citet{liewer2009} for a detailed analysis of the associated filament eruption and CME). They showed that an impulsively generated propagating disturbance has similar kinematics in all four EUVI passbands (304, 171, 195, and 284~\AA), each being inconsistent with the assumption of constant acceleration. In the 304~\AA\ passband, they found that the disturbance showed a velocity peak of $238\pm20$~km~s$^{-1}$  within $\sim$28 minutes of its launch, varying in acceleration from 76 to -102~m~s$^{-2}$ . Comparable velocities and accelerations were found in the coronal 195~\AA\  passband, while lower values were found in the lower cadence 284~\AA\ passband. In the higher cadence 171~\AA\ passband, the velocity varies significantly, peaking at 475$\pm$47~km~s$^{-1}$  within $\sim$20 minutes of launch, with a variation in acceleration from 816 to -413~m~s$^{-2}$ . The high image cadence of the 171~\AA\ passband (2.5 minutes compared to 10 minutes for the similar temperature response 195~\AA\ passband) is found to have a major effect on the measured velocity and acceleration of the pulse, which increase by factors of $\sim$2 and $\sim$10, respectively. These results suggest that previously measured CBF velocity measurements (e.g., using EIT) may actually have been a lower-limit. It should be noted that the near-stationary to rapidly-moving nature of the disturbance in the 171~\AA\ passband led to an apparent rapid rise in velocity soon after its launch. Furthermore, the limited number of data-points in the 304, 195, and 284~\AA\ passbands make it challenging to accurately determine the velocity and acceleration to a high level of accuracy. It is also possible that the numerical differencing scheme used (based on a Lagrange polynomial expansion) may have enhanced a number of initially insignificant trends in the data. That said, the decreasing velocity of CBF was confirmed by \citet{veronig2008}, who also analysed \emph{STEREO} observations of the 19~May~2007 event. They constructed distance-time diagrams of the wave derived by calculating the mean distance of the fronts from the wave center along great circles in the lower corona. A linear fit to the distance-time diagram gave a mean wave velocity of 260~km~s$^{-1}$, comparable to the findings of \citet{long2008}. A quadratic fit yields a start velocity of 460~km~s$^{-1}$ with a constant deceleration of -160~m~s$^{-2}$. The velocity evolution demonstrates that the wave decelerates, with the earliest velocities as high as 400--500~km~s$^{-1}$ . This is somewhat faster than the typical range of velocities reported for CBFs by \citet[][170--350~km~s$^{-1}$]{klassen2000}, which both \citet{veronig2008} and \cite{long2008} attribute to the much better cadence of \emph{STEREO}/EUVI. 

More complex CBF kinematics, including non-moving fronts, have been reported. These are particularly difficult to reconcile with wave-based models. For example, \citet{zhukov2009} reported a nearly symmetric wave-front that exhibited a peculiar velocity profile measured using two independent methods. After a short period propagating at a speed of $\sim$100~km~s$^{-1}$, the wave was then observed to move at a low velocity (around 20--40~km~s$^{-1}$) for about 30 minutes, before being re-accelerated to speeds of around 200~km~s$^{-1}$. This observations makes it difficult to understand how such a velocity change could be accounted for by a freely propagating coronal wave. However, \citet{zhukov2009} suggest that such behavior is possible for erupting prominences, which have previously been observed to exhibit a punctuated kinematic evolution \citep[e.g.,][]{romano2003}. Stationary CBFs have also been observed at coronal hole boundaries \citep[e.g.,][]{delannee2007}. This is discussed further in the section on interactions of CBFs with other coronal structures (Section~\ref{sect:interaction}).

\begin{figure*}[!t]
\begin{center}
\includegraphics[width=0.85\textwidth]{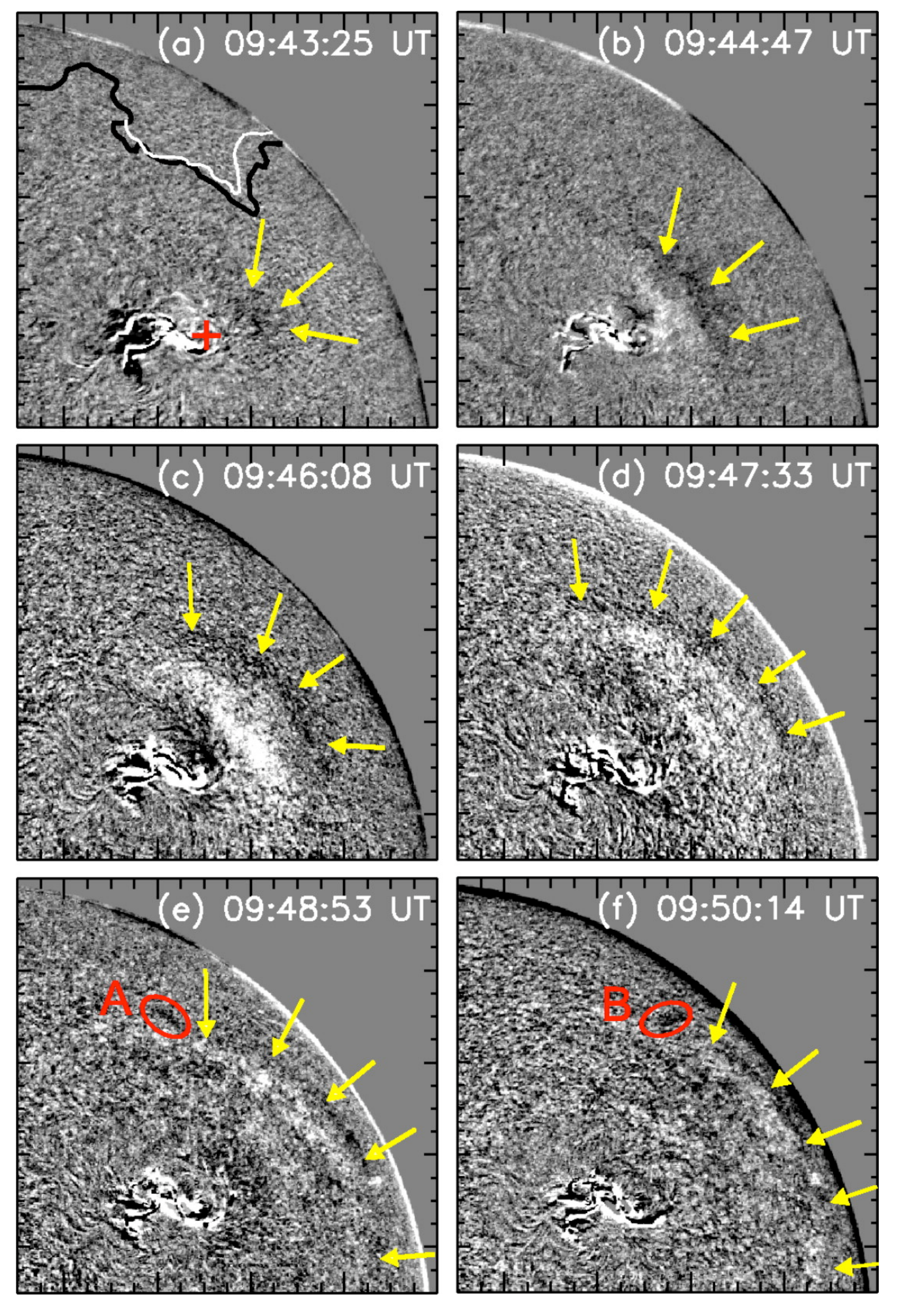}
\caption{Sequence of H$\alpha+0.4$~\AA\ running-difference images. On both axes, the plotted field of view extends from 100 to 1000 arcsec from Sun centre. The leading edges of the Moreton wave fronts (seen as dark fronts in H$\alpha$ red wing observations, with the regions of lower emission shown in white) are indicated by arrows. In panel (a), the determined wave ignition center is indicated by a cross; the coronal hole boundaries are marked by white and black lines for the inner and outer estimates, respectively. In panels (e) and (f), activated features are indicated as ``A" and ``B", respectively \citep{veronig2006}.}
\label{fig:veronig_06}       
\end{center}
\end{figure*}

\section{Chromosphere Signatures of Coronal Bright Fronts}
\label{sect:chrom_sig}

The relationship, both spatially and temporally, between CBFs and other global wave-like phenomena, such as Moreton waves, is difficult to determine, primarily due to the lack of simultaneous observations in multiple wave-bands. \citet{warmuth2001} found compelling evidence that Moreton and CBF travelled co-spatially. This is at odds with the results of \citet{eto2002}, who found that the times of visibility for the Moreton wave did not overlap those of the CBF. In addition, their measurements of the velocity and location of the Moreton wave showed that the Moreton wave and CBF had substantial physical differences, with the Moreton wave preceding the CBF, something which would appear to rule out the identification of a CBF as a fast-mode MHD shock. From a sample of fourteen CBFs, \citet{okamoto2004} found that eleven were associated with filament eruptions, while three were associated with Moreton waves. This provided further evidence that the occurrence of CBF does not necessarily imply the presence of a Moreton wave, suggesting that the formation mechanism may be more complex. In this vein, \citet{bala2007} suggest that Moreton waves are only observed when the velocity of the disturbance exceeds Mach~2. This could explain the observations of CBF without associated Moreton waves. It is clear though that Moreton waves do share some of the characteristics of CBFs, with \citet{veronig2006} reporting observations of the interaction of a Moreton wave and a coronal hole for example. As can be seen in Figure~\ref{fig:veronig_06}, they found that the propagation of the Moreton wave perpendicular to the coronal hole boundary was stopped by the coronal hole, mirroring similar observations of CBFs. Stationary brightenings have also been observed in simultaneously observed CBFs and Moreton waves by \citet{delannee2007}. From an observational perspective, the spatial relationship between CBFs and Moreton waves therefore remains unresolved.

\begin{figure*}[!t]
\begin{center}
\includegraphics[width=1\textwidth]{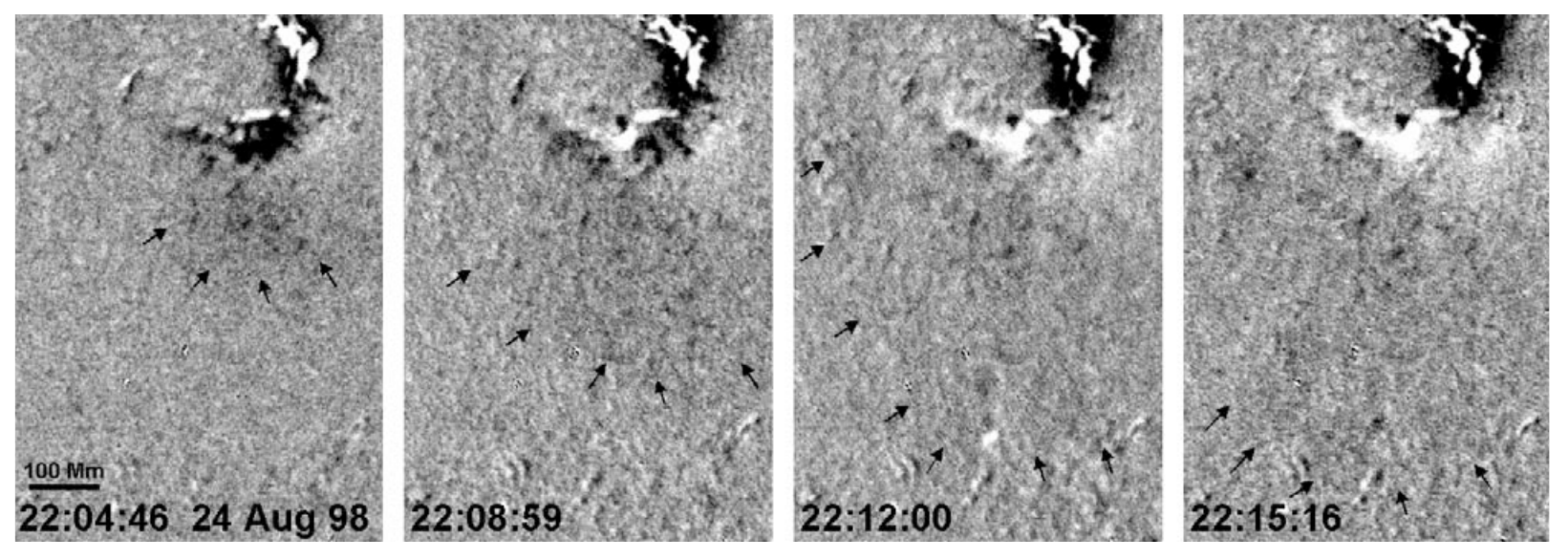}
\caption{Difference images revealing the He~{\sc i} wave of 24~August~1998. The base image was taken at 21:52:50~UT. The vertically oriented bright ribbons in the upper right edge correspond to the main part of the H$\alpha$ flare, whereas the bright horizontally oriented patches below the ribbons are co-spatial with prominent remote brightenings of the H$\alpha$ flare. The He~{\sc i} wave appeared to emanate from this region \citep{vrsnak2002}.}
\label{fig:he1wave}
\end{center}
\end{figure*}

Despite significant research on the relationship between Moreton waves and CBFs, the paucity of observations of Moreton waves over distances comparable with CBFs has made their simultaneous study challenging. To overcome this, \citet{vrsnak2002} studied observations of large-scale disturbances propagating in a passband centred on He~{\sc i} ~(10830~\AA; see Figure~\ref{fig:he1wave}). This passband is useful for examining processes with signatures in the upper chromosphere. Indeed, it was originally proposed by \citet{vrsnak2002} that flare-associated waves observed in He~{\sc i}~(10830~\AA) could be interpreted as the link between H$\alpha$ Moreton waves and CBFs. This was motivated  by observations of the morphology of He~{\sc i} waves, which appeared as an expanding arc of increased absorption roughly corresponding to the H$\alpha$ disturbance, although not as sharply defined. The observed disturbance was also compared to an associated CBF, with areas of reduced He~{\sc i} absorption found to trail behind the front, and thus corresponding to the EUV dimming area.  They also reported that the observed disturbances in near-simultaneous observations in EIT 195~\AA\, He~{\sc i}, and H~{$\alpha$} passbands had closely related kinematical curves, suggesting that they were consequences of a common disturbance. This led \citet{vrsnak2002} to propose that the observations were indicative of a fast-mode MHD coronal shock that was weakly inclined to the solar surface, with the H$\alpha$ front and the He~{\sc i} dip corresponding to the pressure jump in the corona behind the initial pressure front. This interpretation remains controversial.

The usefulness of the He~{\sc i} passband in studying coronal CBFs was underlined by \citet{gilbert2004} who studied two He~{\sc i} wave events with associated CBFs observations in the \emph{SOHO}/EIT~(195~\AA) passband. In both cases, the CBFs and He~{\sc i} waves were observed to travel co-spatially. This suggests that the He~{\sc i} wave is not a product of the same wave producing the CBF, but instead is an imprint of the CBFs in the chromosphere as it propagates through the corona. As noted by \citet{gilbert2004}, if the He~{\sc i} and CBFs were both produced by the same propagating MHD wave, the He~{\sc i} wave would be observed to lag the CBFs due to the lower characteristic velocity in the chromosphere with regard to the corona. \citet{gilbertNholzer2004} attempted to analyse the events considered by \citet{gilbert2004} in greater detail. In particular, they carried out a detailed analysis of the disturbance in He~{\sc i} velocity measurements. The observation of the wave in velocity images was explained as being due to the propagation of slow-mode waves in the chromosphere produced by the coronal fast-mode wave. It was also proposed that the initiation mechanism of these disturbances may be both flare and CME related, with the observation of multiple waves used to support this hypothesis. In this case, it was proposed that the initial pulse was due to the CME, with the following pulses caused by the associated flare. \citet{gilbertNholzer2004} noted that neither of these signatures (in velocity and intensity) are a rare occurrence. 

Although there is evidence to treat He~{\sc i} waves as the missing link between CBFs and Moreton waves, observations of He~{\sc i} waves are scarce, inhibiting any detailed analysis of this phenomenon. In addition, the complex formation mechanism of the He~{\sc i} passband (see the appendix in \citet{gilbert2004} for details) makes it a non-trivial task to identify and physically explain propagating disturbance fronts observed in this passband. However, it must be noted that these observations do have the potential to play an important role in the theory that combines all observations of large-scale propagating disturbances.

\section{EUV and X-ray Signatures of Coronal Bright Fronts}
\label{sect:corona_sig}

Although CBF have primarily been studied in the coronal 195~\AA\ passband, they have also been identified in other bands, such as 171~\AA, 284~\AA, and X-ray. During an extended period of high cadence 284~\AA\ observations, a CBF was identified by \citet{zhukov2004}. This was the first observation of a CBF in this higher temperature passband ($\sim$2~MK), leading these authors to conclude that these phenomena are purely coronal. \citet{long2008} identified a CBF in all four passbands of \emph{STEREO}/EUVI, including the first observations in the waveband centred on 304~\AA\ (see Figure~\ref{fig:longetal}). The 304~\AA\ signature is quite faint and \citet{long2008} concluded that it was due to a contribution from a coronal Si~{\sc xi} (303.32~\AA) line, which has a peak formation temperature of $\sim$1.6~MK, suggesting that the 304~\AA\ passband feature is coronal rather than chromospheric. A more detailed discussion of the EUV spectrum in this wavelength range and its variation from region to region in the solar atmosphere can be found in \citet{brosius1996}. \citet{patsourakos2009} used observations from different passbands from two perspectives to show that the disturbance in the EUV is most apparent in the 1--2~MK temperature range (corresponding to the 171, 195, and 284~\AA\ passbands). \emph{STEREO} quadrature observations were used by \citet[][see Figure~\ref{fig:kienreich_09}]{kienreich2009}  to show that CBFs propagate in a layer of the atmosphere located at $\sim$80--100~Mm above the photosphere. This is similar to the height  derived using triangulation techniques by \citet{patsourakos2009}, who showed that the wave propagates in a layer 90$\pm$7~Mm above the surface. These heights are comparable to a coronal scale height of 50--100 Mm for quiet-Sun temperatures of 1--2~MK. 

\begin{figure*}[!t]
\begin{center}
\includegraphics[width=1\textwidth]{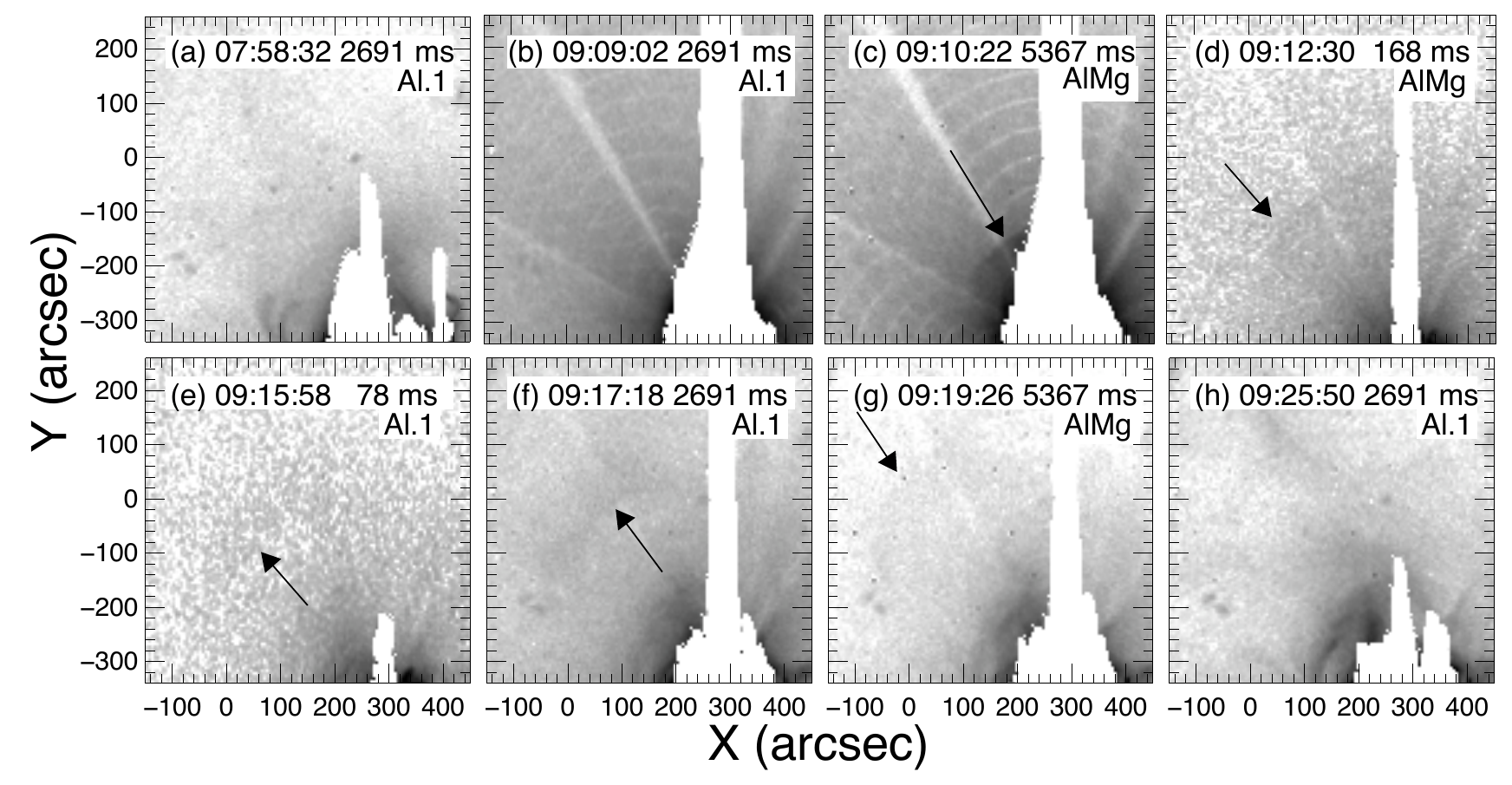}
\caption{X-ray wave observed by \emph{Yohkoh} and reported by \citet{khan2002}. The wave is identified by the black arrow. The white region in the centre of each image and the circular and radial lines are instrument effects due to saturation from the flare.}
\label{fig:xraywave}
\end{center}
\end{figure*}

Propagating disturbances similar to ``EIT waves" were first observed in soft X-ray emission from the solar corona using data from the \emph{Yohkoh} spacecraft \citep{khan2002}. Although \emph{Yohkoh} was programmed to switch to high cadence, small field-of-view images upon the initiation of a flare, in this case \emph{Yohkoh} was passing through the South Atlantic Anomaly at the time of the event, with the result that flare mode was not triggered, allowing the effect of the flare on the full-disk to be identified (see Figure~\ref{fig:xraywave}). In this example, the wave is difficult to identify, but it was possible to obtain an average speed between the two clearest leading edges of $546\pm21$~km~s$^{-1}$,  assuming that the leading edge was moving with constant velocity. With regard to the timing of the X-ray wave, a Moreton wave was first observed $\sim$72 seconds prior to the first X-ray wave observation, a CBF was first observed between the two X-ray wave sightings. The extrapolated leading edge of the X-ray wave was observed to closely match the Moreton wave and CBF, suggesting that all three waves must be closely related to each other. Based on these observations, it was proposed that the observed X-ray wave was a blast wave associated with either chromospheric evaporation or an erupting structure slightly offset from the flaring region.

The next observation of an X-ray wave was made by \citet{hudson2003}, with an X-ray wave associated with an X-class flare and Type~{\sc ii}, {\sc iii}, and {\sc iv} radio bursts. In this event, a large loop structure to the north of the flare core is observed to explode outwards, with the wave appearing to emanate from this location. The event was associated with a very strong intensity enhancement, with the electron temperature in the wave estimated to be $\sim$2--4~MK, and the Mach number of the shock estimated to be $\sim$1.1--1.3 (these estimates are dependent on the assumed coronal ambient temperature). \citet{hudson2003} concluded that the observations were consistent with the interpretation of the X-ray emission front as the signature of a coronal shock wave at a low Mach number. The cause of the shock wave was hypothesized to be due to plasma motions that began at a relatively small distance from the flare core, with this conclusion due to the close association between the excitation of the wave and the beginning of the impulsive phase of the flare. The problems associated with low cadence, partial field-of-view observations using \emph{Yohkoh} were overcome by \citet{warmuth2005}, who used higher cadence \emph{GOES}/SXI images. This data was combined with EUV, H$\alpha$ and He~{\sc i} data for six events, with the conclusion being that all observations could be explained by a single decelerating large-amplitude wave or shock.

The most recent X-ray wave observations have been made by \citet{attrill2009} using data from  \emph{Hinode}/X-Ray Telescope \citep[XRT;][]{golub2007}. By combining data from both \emph{Hinode}/XRT and \emph{STEREO}/EUVI (171, 195, and 284~\AA\ passbands) they found the strongest components of the observed disturbance to be largely co-spatial in all observed bandpasses, suggesting that all observations correspond to the same disturbance. In addition, \emph{STEREO}/COR1 images were used to show that the core coronal dimmings in both EUVI and XRT map to the core of the CME, with the secondary coronal dimmings mapping to the CME cavity and the diffuse coronal wave mapping to the outermost edge of the expanding CME shell. The suggestion made by \citet{attrill2009} was that the observed coronal ``wave" is actually the erupting CME. While the erupting CME may have shocked the plasma enabling X-ray emission, the observations appear to rule out a wave origin for the observed X-ray disturbance.

Spectroscopic observations of CBFs have been limited due to the small field-of-view and low rastering cadence of EUV spectrometers. Using spectral data from the Coronal Diagnostic Spectrometer \citep[CDS;][]{harrison1995} instrument on \emph{SOHO}, \citet{harraNsterling2003} observed a coronal wave feature  associated with a solar eruption and flare on 1998~June~13 \citep[cf.][]{willsdaveyNthompson1999}. They pointed out some aspects of this event that were not covered by \citet{willsdaveyNthompson1999}. For example, \citet{harraNsterling2003}  found that the brightest part of the wave front separated in two, with a weak part of the wave running ahead of the brightest wave front. The weak wave passes through the CDS field-of-view but shows velocities less than $\sim$10~km~s$^{-1}$. They concluded that their observations support the work of  \citet{chen2002}, who suggest that coronal waves consist of a faster propagating, piston-driven portion and a more slowly propagating portion due to the opening of field lines associated with an erupting filament. Using spectra from the EUV Imaging Spectrometer \citep[EIS;][]{culhane2007} on \emph{Hinode},  \citet{asai2008} found blue-shifts associated with a CBF launched during a CME eruption. The center of the blue-shifted component had a Doppler velocity of $\sim$100~km~s$^{-1}$, while a velocity of 450~km~s$^{-1}$ was observed along the slit. These components were only observed in the hottest lines of the raster (Fe~{\sc xv} and Ca~{\sc xvii}), which have formation temperatures of $\sim$2~MK. In \emph{Hinode}/XRT images, these blue-shifted components were found to correspond to a coronal wave-like feature. The authors suggested that these observations provide evidence for the existence of fast-mode MHD shock waves.

The small number of papers on the multi-thermal structure of CBFs highlights that detections of X-ray waves are rare, and hence, the basic physics remains ill-defined. It is therefore difficult to make a detailed comparison of observation and theory due to the small sample sets available and the nature of the observations. These problems can only be overcome through the use of high cadence coordinated EUV and X-ray observations such  as can be provided by \emph{Hinode}, \emph{STEREO}, and \emph{SDO}.

\section{Coronal Bright Fronts and Radio Emission}
\label{sect:radio}

The primary radio bursts associated with coronal waves are metric Type~{\sc ii} bursts. It has been noted that the majority of flares of any magnitude range are not necessarily accompanied by Type~{\sc ii}  activity, implying that an additional condition must be present for the bursts to form \citep{roberts1959}.  With the discovery of Moreton waves \citep{moreton1960}, it was noticed that there was a strong correlation between metric Type~{\sc ii} radio bursts and Moreton waves \citep{kai1970}. Further investigation indicated that $\sim$70~\% of observed flare (Moreton) waves were associated with metric Type~{\sc ii} bursts \citep{pinter1977}. In addition, the estimated speed of the radio observations exceeded, but were typically proportional to the associated Moreton wave velocities. 

With the launch of \emph{SOHO}, $\sim$90\% of Type~{\sc ii} bursts were found to be associated with CBF transients \citep{klassen2000}.  It was noted that the correlation between metric Type~{\sc ii} bursts and CMEs increased dramatically as the origin of radio activity approached the solar limb, with an additional positive correlation between CME visibilty and Type~{\sc ii} bursts \citep[see also][] {biesecker2002}. The correlation was also high for CMEs with a velocity exceeding $\sim$400~km~s$^{-1}$ \citep{cliver1999}. Type~{\sc ii} radio bursts are signatures of shock waves travelling outwards in the upper corona. As the mean CBF velocity of 290~km~s$^{-1}$ is well above the sound speed in the corona, these waves were considered by many authors as fast magnetosonic waves propagating nearly perpendicular to the ambient magnetic field in the low corona \citep{mann1999,klassen2000}. 

\begin{figure}[t!]
\begin{center}
\includegraphics[keepaspectratio, width=0.8\textwidth,trim = 0mm 0mm 0mm 0mm,clip=]{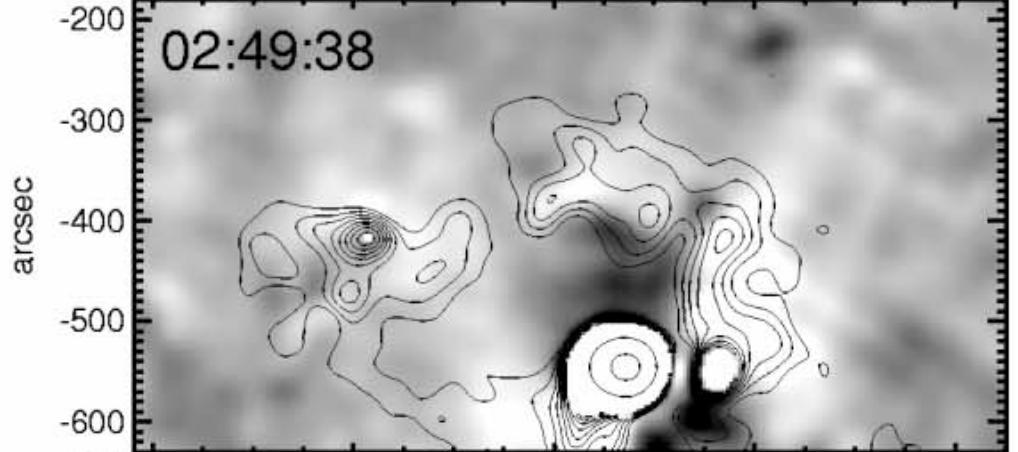}
\includegraphics[keepaspectratio, width=0.8\textwidth,trim = 0mm 0mm 0mm 0mm,clip=]{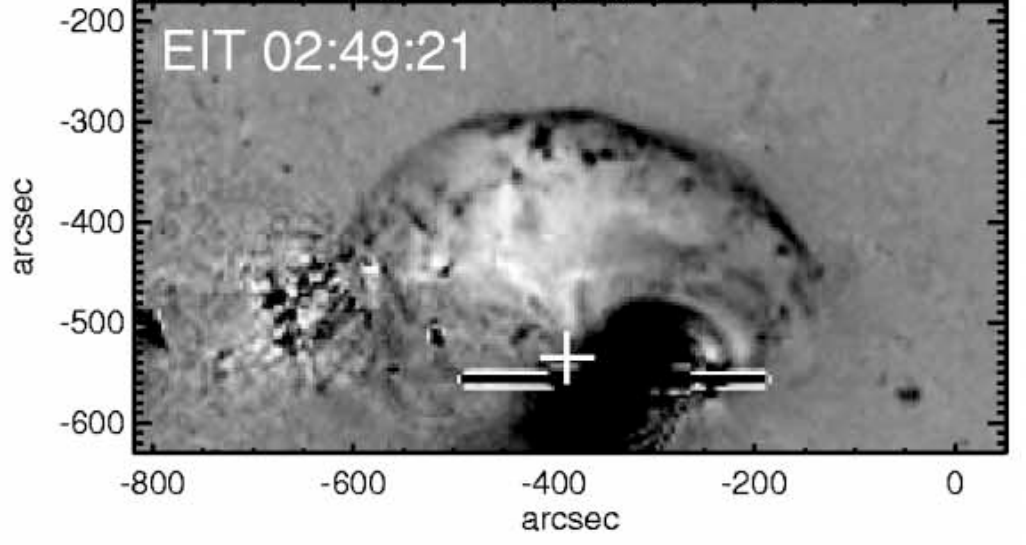}
\caption{NoRH 17~GHz observations of the propagating disturbance on 24~September~1997 \citep{white2005}. The top panel shows a radio image obtained at 02:49:38~UT over-plotted with contours at brightness temperature multiples of 600~K up to 7200~K, the top three contours are at 0.365, 1.96, and 4.85 $\times10^6$~K. The bottom panel shows an EIT difference image from 02:49:21~UT (Note: Black represents regions of higher emission).}
\label{fig:white_2005}
\end{center}
\end{figure}

\citet{warmuth2004b} studied 12 flare-associated wave events in order to determine their physical nature, using H$\alpha$, EUV, Helium~{\sc i}, SXR and 17~GHz radioheliographic data. The wavefronts observed in the various spectral bands were found to follow kinematical curves that were closely related, implying that they were signatures of the same physical disturbance. In particular, \citet{warmuth2004b} found evidence for deceleration, perturbation profile broadening, and perturbation amplitude decrease. They concluded that the pulses could be interpreted as freely propagating fast-mode MHD shock formed from a large-amplitude disturbance of the medium. It was shown that this scenario could account for a large majority of the observed properties of the waves in the various spectral bands, as well as for the associated metric Type~{\sc ii} radio bursts.

\citet{white2005} studied sensitive Nobeyama Radioheliograph (NoRH) observations of the 1997~September~24 CBF (see Figure~\ref{fig:white_2005}), finding a propagation speed of 830~km~s$^{-1}$. This led them to suggest that the EIT instrument cadence may have the effect of producing underestimates of disturbance velocities by a factor of 2-3 \citep[cf.][]{long2008}. The radio spectrum, which was poorly constrained using only 17 and 34~GHz observations, was judged to be consistent with optically thin coronal emission rather than chromospheric emission. In contrast to the findings of \cite{warmuth2004b}, \citet{white2005} found no evidence for deceleration in 17~GHz radio images of the event (it should be noted that the signal-to-noise of the 34~GHz images was too low for the wave to be adequately identified). 

\citet{vrsnak2005} studied broadband Nan\c{c}ay Multifrequency Radioheliograph observations associated with a Moreton/CBF wave. They found radio emission in the range 151--327~MHz that was considerably weaker than the flare-related Type~{\sc iv} burst. The emission centroid propagated at a height of 0-200~Mm above the solar limb and was intensified when the disturbance passed over enhanced coronal structures. A quadratic fit to the metre-wave observations showed continuous deceleration of the Moreton wave, CBF and radio wave (the deceleration of the CBF was -310$\pm$50~m~s$^{-2}$, and for the NRH wave -350$\pm$50~m~s$^{-2}$). Consistent with \citet{white2005}, they proposed that the NRH wave signature is optically thin gyrosynchrotron emission excited by the passage of the coronal MHD fast-mode shock. 

It is evident that from the limited number of papers that have considered radio emission associated with CBF that there are many unanswered questions regarding their physical connection. At metre wavelengths it is not yet clear if both CBF and Type~{\sc ii} bursts are excited by the same coronal shock wave. In contrast, at microwave frequencies the limited number of observations and lack of high resolution spectra make it difficult to determine the physical mechanism responsible for radio-wave excitation. The Frequency Agile Solar Radiotelescope \citep[FASR;][]{bastian2004} in Owens Valley, California will be sensitive to emissions from chromospheric to coronal heights, and will provide a complete view of chromospheric and coronal waves, dimmings, and the interaction of waves with surrounding structures such as active regions \citep[e.g.,][]{ofmanNthompson2002}.

\section{Interaction with Coronal Structures}
\label{sect:interaction}

Observations of the interaction of CBF with active regions and coronal holes are rare, mainly as a result of the low 12 minute cadence of \emph{SOHO}/EIT. However, a number of authors have shown that CBFs tend to avoid active regions \citep{thompson1999,willsdaveyNthompson1999} and stop at the boundaries of coronal holes \citep{thompson1998,tripathi2007} as well as near the separatrix between active regions, where they may appear as a stationary front \citep{delannee1999,chenNfang2005}. These observations were reproduced in numerical simulations treating CBF as fast-mode MHD waves \citep{wang2000,wu2001,ofmanNthompson2002}, showing that the wave undergoes strong refraction and deflection when interacting with active regions and coronal holes, which are both high Alfv\'{e}n velocity regions compared to the quiet Sun. 

Using $\sim$1 minute cadence \emph{TRACE} observations, \citet{ballai2005} showed that CBFs have a well defined period and energy.  A wavelet analysis indicated a very strong signal with period of approximately 400~s, which the authors concluded is strong evidence for the periodic wave nature of CBFs.  From the excitation of transverse coronal loop oscillations by a CBF, and supposing that the entire energy of CBFs has been transmitted to the oscillating loop, \citet{ballai2005} estimated the minimum energy of a CBF using
\begin{equation}
E = \frac{\pi L (\rho_i R^2 + \rho_e / \lambda_e^2)}{2}\left(\frac{x_{max}-x}{t_{max}-t}\right)^2
\end{equation}
where $L$ is the length of the loop, $R$ is its radius, $\rho_i$ and $\rho_e$ are the densities inside and outside the loop, $x_{max}$ is the maximum deflection of the tube that occurs at $t_{max}$, and $x$ is an intermediate deflection that occurs at a time $t$. For typical coronal values, this gives a minimum energy of an CBF to be $\sim$3.4$\times$10$^{18}$~J, comparable to the energy of a nanoflare.

The absence of high-cadence full-disk EUV observations was overcome by \citet{veronig2006} who used H$\alpha$ images to supplement EUV observations of a propagating disturbance, and to study the interaction of a Moreton/CBF wave with a coronal hole. The large angular extent of the Moreton wave allowed them to study the wave kinematics in different propagation directions with respect to the location of a polar coronal hole. In particular, they found that the wave segment whose propagation direction is perpendicular to the coronal hole boundary was stopped, which is in accordance with observations reported from CBFs \citep[e.g.,][]{thompson1999}. Interestingly, \citet{veronig2006} found that the wave signatures can be observed up to 100~Mm inside the coronal hole, where the front orientation is perpendicular to the coronal hole boundary.

\begin{figure}[t!]
\begin{center}
\includegraphics[keepaspectratio, width=0.45\textwidth,trim = 0mm 45mm 0mm 40mm,clip=]{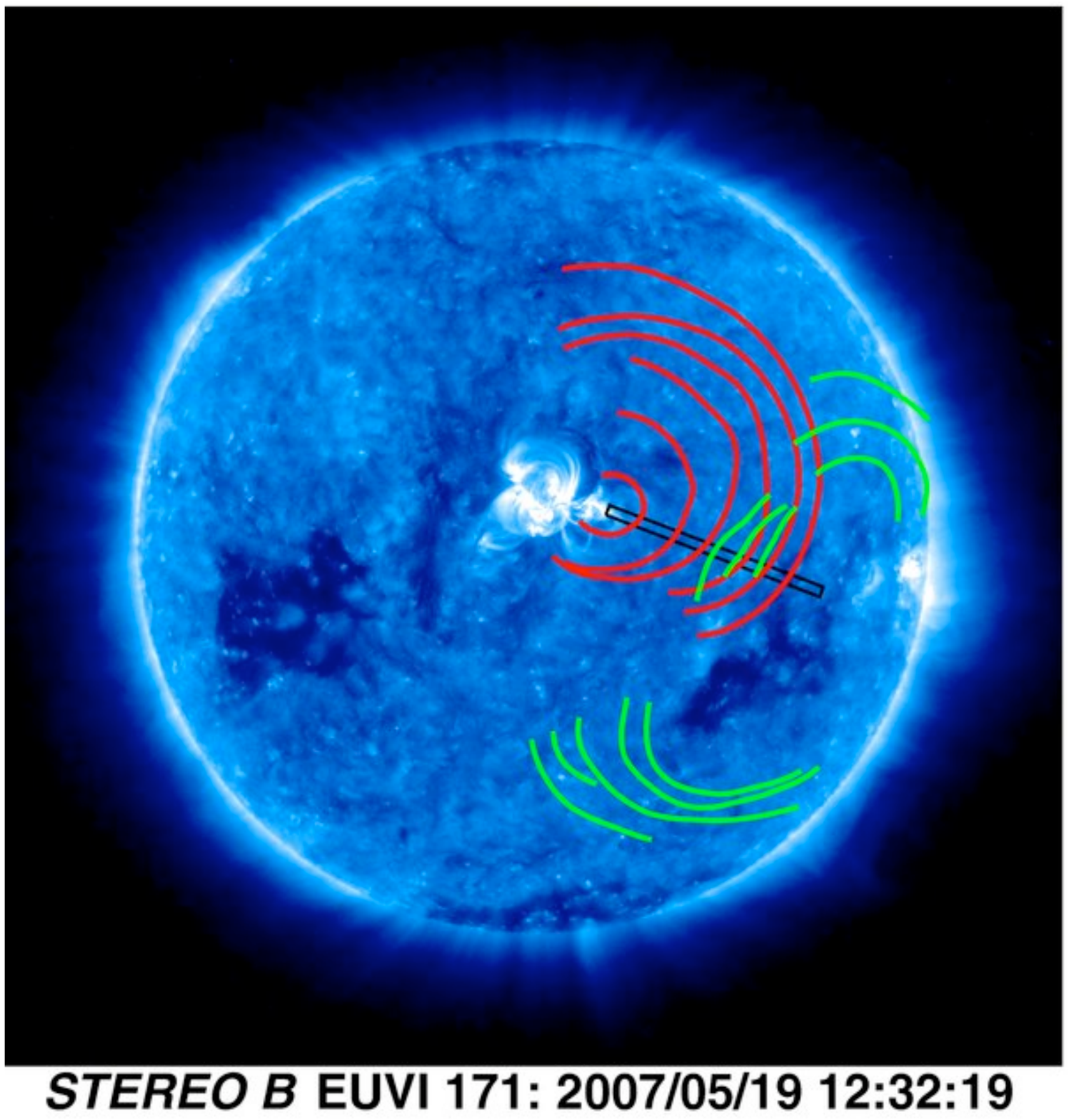}
\includegraphics[keepaspectratio, width=0.48\textwidth,trim = 0mm 50mm 0mm 50mm,clip=]{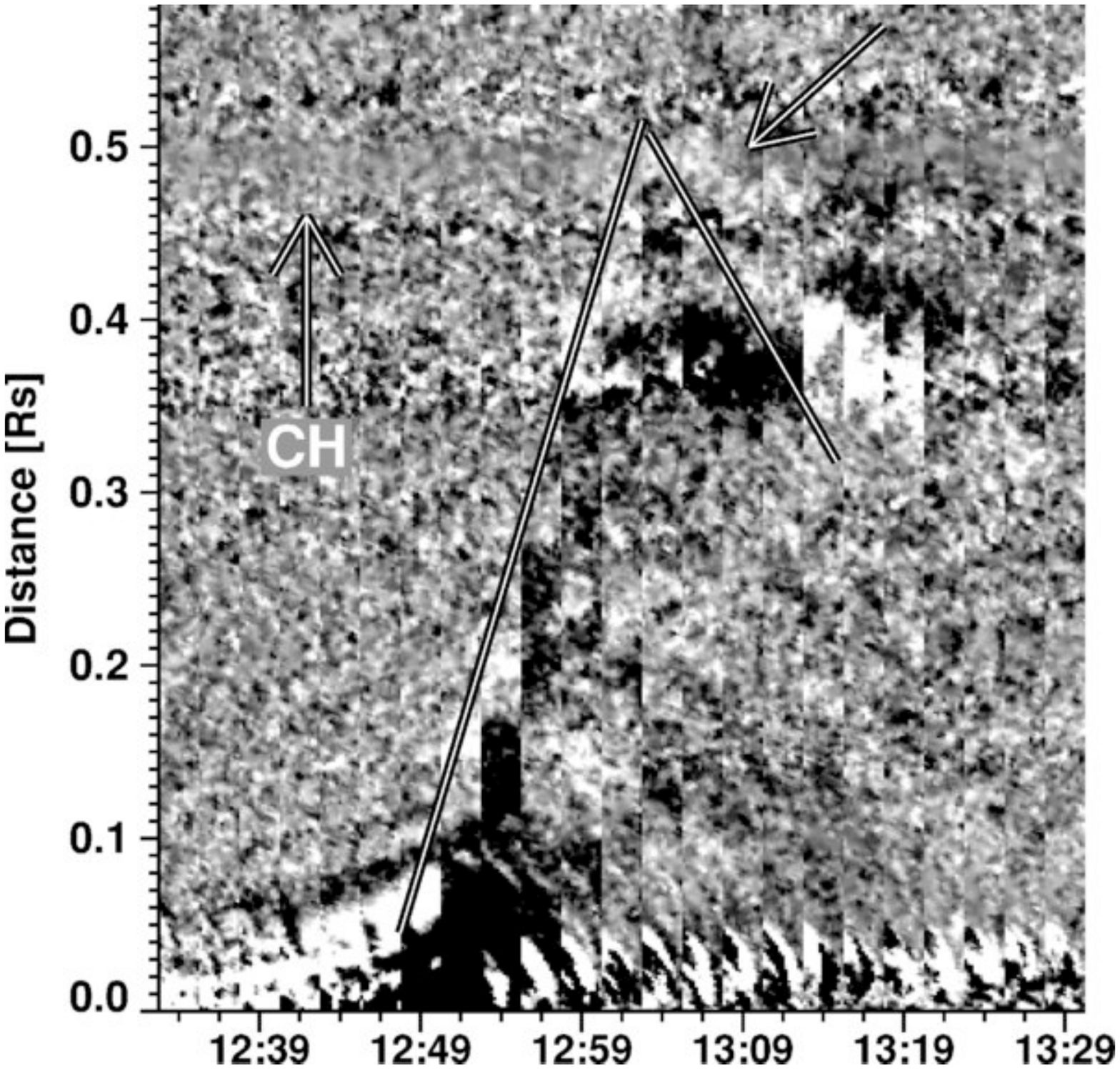}
\caption{{\it Left:} Sketch of the EUV wave at different times superposed on a \emph{STEREO}/EUVI 171~\AA\ image from \citet{gopalswamy2009}. Direct wave (red) and reflected waves (green) are distinguished. A rectangular slit, shown in black, was used to investigate the wave motion shown on the right. {\it Right:} Distance-time plot made by stacking EUV difference images within the rectangular slit at different times. Note the reflection around 13:00 UT (indicated by the arrow).}
\label{fig:gopal_2009}
\end{center}
\end{figure}

The first CBF studied using \emph{STEREO}/EUVI allowed much higher cadence observations to be obtained (2.5 minutes in the 171~\AA\ passband). Both \citet{long2008} and \citet{veronig2008} noted that the CBF showed strong evidence of reflection from a coronal hole to the southwest of the active region origin, with some evidence of refraction also apparent. \citet{gopalswamy2009} further studied this event, focussing on the interaction of the wave with this nearby coronal hole and its apparent reflection (see Figure~\ref{fig:gopal_2009}). In particular, they found that the reflected wave speed was significantly different than the incident wave speed. While the incident wave had a velocity of $\sim$384~km~s$^{-1}$, the reflected velocity varied from $\sim$200--600~km~s$^{-1}$ depending on direction of propagation. \citet{gopalswamy2009} concluded that EUV disturbances are waves driven by a CME flux rope rather than brightening resulting from non-wave processes suggested in the literature \citep[e.g.,][]{delannee1999,attrill2007}. Furthermore, \citet{gopalswamy2009} stated that an associated metric Type~{\sc ii} burst was clear evidence of a fast mode shock, which must have formed due to the steepening of the magnetosonic wave.

Theoretically, \citet{wang2000} determined the distribution of the magnetosonic velocity $v_f$ in the corona using a current-free extrapolation of the measured photospheric field and a density scaling law for coronal loops. In agreement with observations, they found that the waves are deflected away from active regions and coronal holes, where $v_f$ is large and that they are refracted upward as they propagate away from their initiation point, since $v_f$ falls off rapidly above active regions. \citet{ofmanNthompson2002} studied the interaction of CBFs with coronal active regions using a three-dimensional MHD model. The active region was modelled by an initially force-free, bipolar magnetic configuration with gravitationally stratified density. The CBF was launched at the boundary of the region, as a short time velocity pulse that travels with the local fast magnetosonic speed toward the active region. They found that the CBF undergoes strong reflection and refraction, in agreement with observations.

\section{Relationship of Coronal Bright Fronts to Flares and CMEs}
\label{sect:flare_cme}

It is now clear that CBF are strongly associated with CMEs, but the exact physical mechanism by which they are related remains unclear. Although the connection between these two phenomena was originally noted by \citet{thompson1998} and \citet{thompson1999}, there has been much debate as to how the morphological and kinematic properties of CBFs can be explained within the context of CMEs evolution, or if they can be excited impulsively by a solar flare \citep[e.g.,][]{warmuth2001}.

\citet{biesecker2002} analysed the statistics of a catalogue of events from \citet{thompson2009} by assigning a quality rating to each CBF and then correcting for other observational biases. As might be expected, a poor correspondence was found between X-ray flares and CBFs occurring at the limb. For disk events, a high rate of coincidence was found, but far from a perfect correspondence. Only for the highest quality of waves was there a one-to-one correlation. They reported that there was no evidence for an impulsive increase in the GOES X-ray intensity at the time of a CBF, although they did show that there was a tendency for waves with a higher rating to be associated with larger flares. Regarding CMEs, \citet{biesecker2002} found that there was a poor correspondence between CBFs occurring within 60 degrees of the central meridian and CMEs. However, when only CBFs within 30 degrees of the limb were examined, a dramatic correlation between the occurrence of CMEs and CBFs was found. In fact, \citet{biesecker2002} conclude that  ``If an EIT wave is observed, there must be a CME; however, the converse is not necessarily true". This study raised serious questions for the shock or blast wave interpretation of CBFs. More recently, these results were supported by the statistical analysis of a smaller data-set by \citet{chen2006}. In order to determine whether CBFs are generated by CMEs or pressure pulses in solar flares, they studied fourteen non-CME-associated energetic flares. They found that none of the flares in their sample were associated with CBFs. Their analysis indicates that CBFs and expanding dimmings appear only when CMEs are present. \citet{chen2006} thus concluded that it is unlikely that pressure pulses from flares generate CBFs. Similar results were reported by \citet{chen2009} using data from \emph{SOHO}/EIT and the Mauna Loa Mk-III Coronameter. Based on this analysis, they suggested that CBFs/dimmings are the EUV counterparts of the CME leading cavity.

\begin{figure*}[!t]
\begin{center}
\includegraphics[width=1\textwidth,trim = 0mm 60mm 0mm 55mm,clip=]{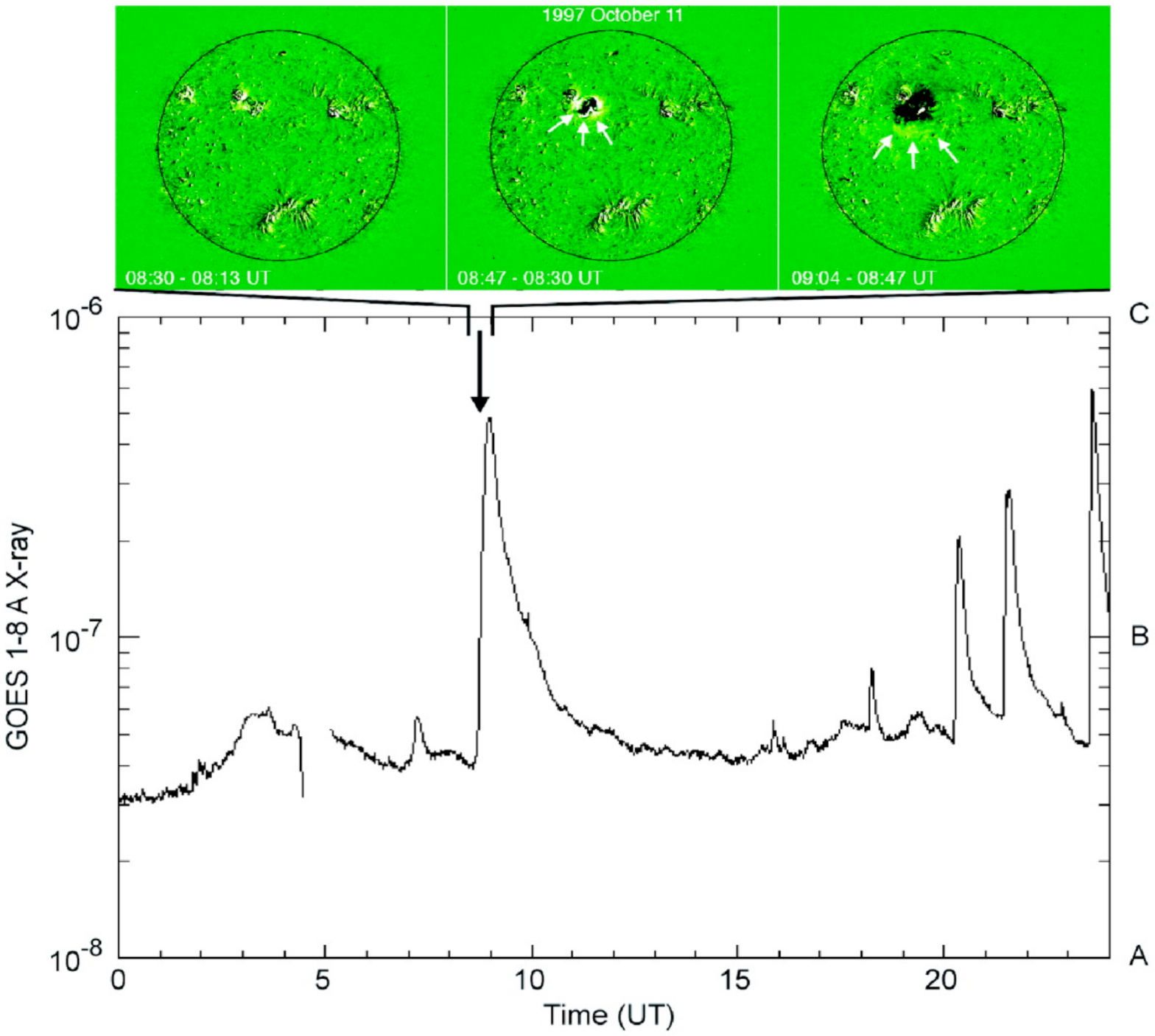}
\caption{{\it Top:} EIT running difference images showing the development of the wave on 11~October~1997 as analysed by \citet{cliver2005}. White arrows indicate the leading edge of the wave. {\it Bottom:} GOES 1--8~\AA\ plot for 11~October~1997, with the B4 soft X-ray flare associated with the wave indicated by an arrow. ``A", ``B", and ``C" in the right-hand margin indicate GOES peak soft X-ray classes.}
\end{center}
\label{fig:cliver_05}       
\end{figure*}

An additional large-scale statistical study of CBFs was carried out by \citet{cliver2005}. They found that approximately half of the large-scale coronal waves identified in images obtained by \emph{SOHO}/EIT were associated with small solar flares with soft X-ray intensities below C--class (see Figure~\ref{fig:cliver_05} for example). Their results indicate the need for a special condition that distinguishes flares with CBFs from the vast majority of flares that lack wave association. While it was thought that this special condition must be the presence of an associated CME, \citet{cliver2005} suggest that this is not sufficient for a detectable CBF. This can be explained by the fact that $\sim$5 times as many front-side CMEs as CBFs occurred during their study period. \citet{cliver2005} also showed for the first time that CBF association increases with CME speed and width. That is, fast, wide CMEs are more likely to have associated CBFs.

\begin{figure*}[!t]
\includegraphics[width=1\textwidth]{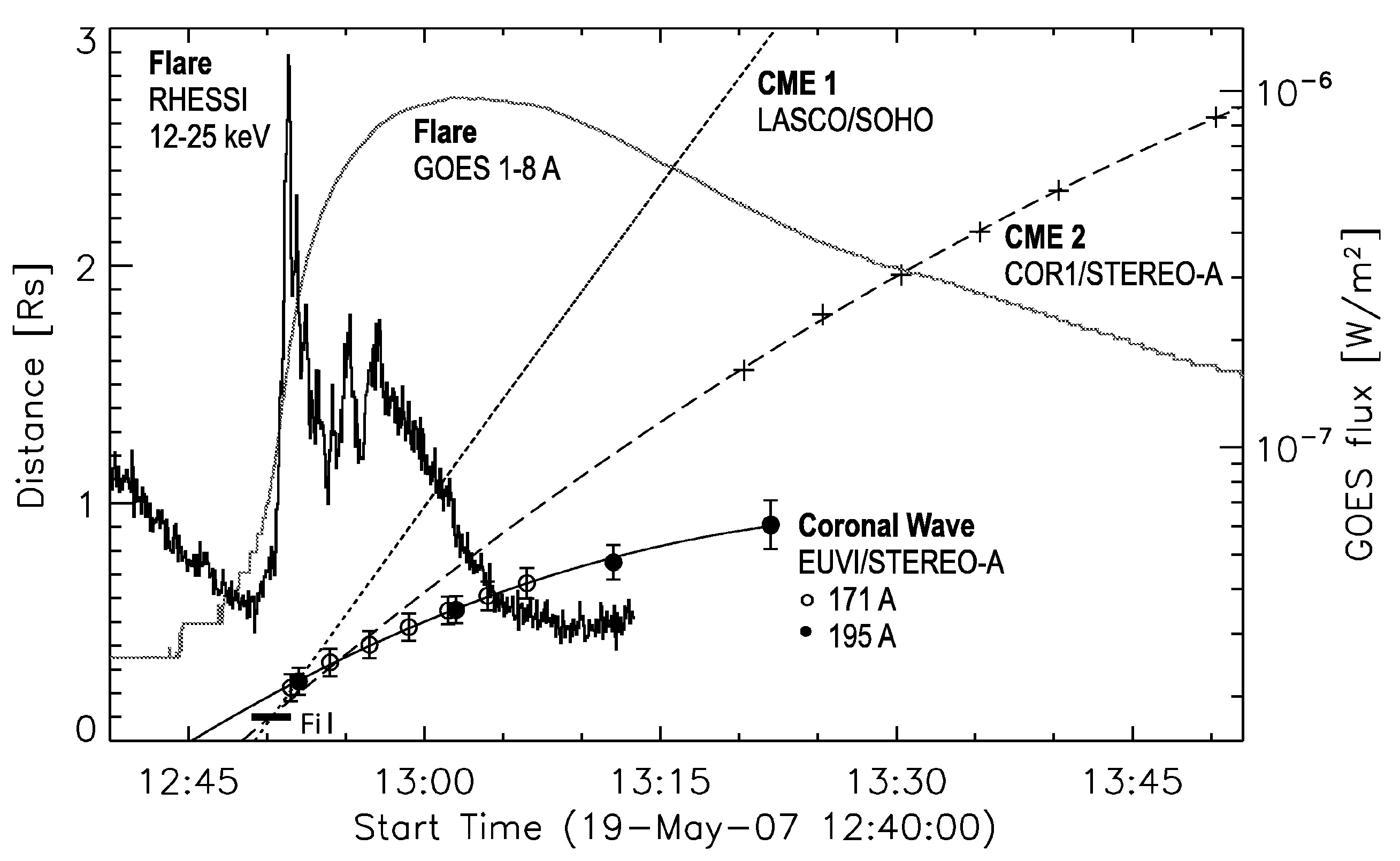}
\caption{Summary plot of the coronal wave kinematics (circles, measured distances; quadratic fit, full lines) together with the flare evolution (grey curve, GOES 1-8~\AA\ soft X-ray flux; black spiky curve, \emph{RHESSI} 12-25 keV hard X-rays) and kinematics of CME~2 observed in \emph{STEREO}/COR1 (plus signs; together with quadratic fit) and the back-extrapolated quadratic fit of CME~1 observed by \emph{SOHO}/LASCO (dotted curve). The horizontal bar indicates the start of the fast filament eruption observed by EUVI. Figure from \citet{veronig2008}.}
\label{fig:veronig_08}       
\end{figure*}

\citet{veronig2008} used analysis of an CBF event with an associated flare and erupting filament/CME hinted at wave initiation by the CME expanding flanks, which drive the wave only over a limited distance. They showed that the associated flare is very weak and occurs too late to account for the wave initiation. They also showed that the kinematics of the coronal wave are quite different from the kinematics of the CME leading edge; the wave is slower than both CMEs and significantly decelerates. Figure~\ref{fig:veronig_08} from \citet{veronig2008} shows a summary plot consisting of: (i) the distance-time diagram of the coronal wave observed by \emph{STEREO}-A/EUVI; (ii) the back-extrapolated (quadratic fit) distance-time diagram of CME~1 observed with \emph{SOHO}/LASCO; (iii) the distance-time diagram of CME~2 observed with \emph{STEREO}-A/COR1; (iv) the flare hard X-ray flux recorded by \emph{RHESSI}; and (v) the flare soft X-ray flux recorded by GOES. From the quadratic fit to the EUVI wave kinematics, they estimated the wave's launch time at $\sim$12:45~UT. The flare 12--25~keV hard X-ray flux starts rising at 12:50~UT with the first and highest peak at 12:51:30~UT. At this time, they had already observed the first EUVI wave front. Such timing argues against a flare-origin of the wave, since the wave needs time to build up a large amplitude or shock to be observable. On the other hand, timing and direction of the erupting filaments indicate that the wave was closely associated with the fast CME~1, since filament 1 disappeared from the H$\alpha$ filter at 12:46 UT, whereas filament 2 remained visible until 12:55~UT.

\section{Physical Interpretations of Coronal Bright Fronts}
\label{sect:modelling}

The physical nature of CBFs remains the subject of continuing debate. One group of theories describes them as fast-mode magnetosonic waves propagating freely in the corona, while the other postulates that they are a signature of magnetic field restructuring during CME launch. When CBFs were first studied  by \citet{thompson1998}, they were thought to be the coronal analogue of Moreton waves, and consequently, were interpreted as being fast-mode MHD waves similar to those suggested by \citet{uchida1968}. It soon became evident that CBFs were closely related to CMEs rather than flares. This lead to a second interpretation of CBFs being the result of expanding CME flanks which re-organised the coronal magnetic field as they swept through the low corona. Both wave and CME-associated models of CBFs are discussed below. 

Waves are predicted to propagate in the solar corona subject to the restoring forces of the magnetic field and gas pressure. In the case of a non-magnetised gas, waves propagate at the local sound speed, given by 
\begin{equation}
c_s = \sqrt{\frac{\gamma p}{\rho}}
\end{equation}
where $\gamma$ is the ratio of the specific heats and $p$ and $\rho$ are the unperturbed gas pressure and density respectively. The typical sound speed for the corona is $\sim$100--200~km~s$^{-1}$. Alternatively, a pure Alfv\'{e}n wave, propagating along a magnetic field line, has a velocity
\begin{equation}
v_A = \frac{B}{\sqrt{4\pi\rho}}
\end{equation}
where $B$ is the magnetic field strength in Gauss. This is typically up to 1000~km~s$^{-1}$ for the corona. A third class, magnetosonic waves, arise from a consideration of both gas and magnetic forces in combination. These waves propagate with velocities given by 
\begin{equation}
v_{f,s}^2 = \frac{1}{2}\left((c_{s}^{2} + v_{A}^{2}) \pm \sqrt{c_{s}^{4} + v_{A}^{4} - 2c_{s}^{2}v_{A}^{2}\cos 2\theta}\right)
\end{equation}
where $\theta$ is the inclination of the wave propagation vector to the magnetic field. This equation has two distinct solutions, with the positive sign corresponding to a fast-mode MHD wave, while the negative sign corresponds to a slow-mode MHD wave. In the case of  $\theta = \pi / 2$ (i.e., the case when the wave is propagating perpendicular to the magnetic field), we obtain
\begin{equation}
v_{f}^{2} = c_{s}^{2} + v_{A}^{2} \hspace{3mm} \textrm{and} \hspace{3mm} v_{s} = 0
\end{equation}
\noindent
for the fast-mode and slow-mode MHD waves. It should be noted that Alfv\'{e}n waves cannot produce the necessary compression to be seen as a brightness enhancement. Slow-mode magnetoacoustic waves are compressive, but their propagation is limited by magnetic field direction; the slow-mode velocity vanishes for propagation perpendicular to field lines. Due to the fact that CBFs propagate at right angles to the Sun's primarily radial magnetic field, they were initially interpreted as fast mode MHD waves \citep[e.g.,][]{thompson2000}.

\citet{wang2000} were the first to simulate CBFs as fast-mode MHD waves. The distribution of the magnetosonic velocity in the corona was determined using a current-free extrapolation of the measured photospheric field and a density scaling law for coronal loops. The average surface-projected expansion speeds were found to be of order 200~km~s$^{-1}$. Their model was unable to account for velocities in excess of 600~km~s$^{-1}$, typical of Moreton waves and Type~{\sc ii} radio bursts, unless it was assumed that the initial disturbance took the form of a strong, super-Alfv\'{e}nic shock. In agreement with observations, \citet{wang2000} showed that fast-mode waves are deflected away from active regions and coronal holes, where the Alfv\'{e}n speed is large. They also found that the waves are refracted upward as they propagate away from their initiation point as a result of the Alfv\'{e}n speed falling off rapidly above active regions.

\begin{figure*}[!t]
\begin{center}
\includegraphics[width=0.8\textwidth]{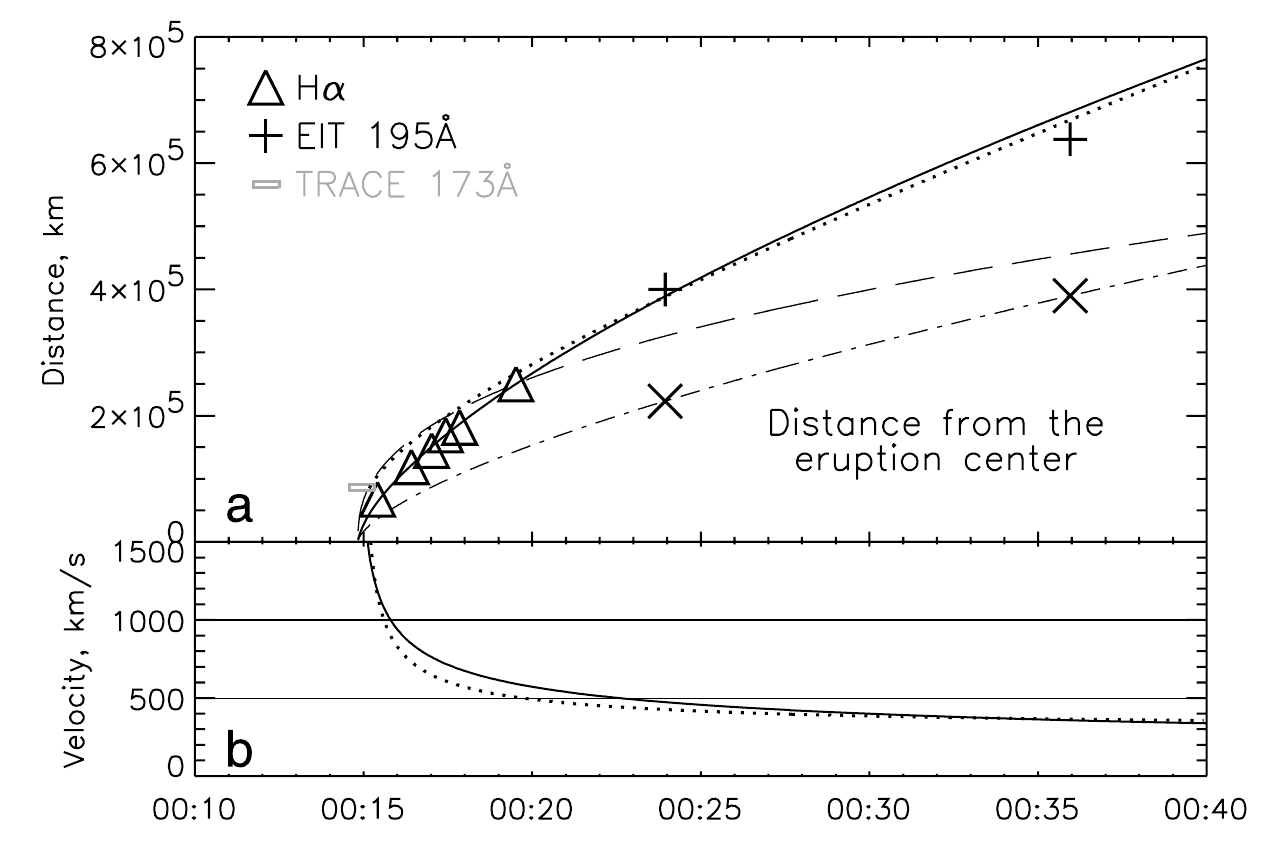}
\caption{(a) Distance-time plots of the Moreton waves (triangles) and CBFs (crosses) measured in the same direction from the eruption center by \citet{grechnev2008}. Rectangles denote the first manifestation of a wave in a \emph{TRACE} 173~\AA\ image. The curves show a strong shock with density distributions of $\rho$ = const (dashed) and $\rho \propto r^{-2}$ (solid) and a weak shock with $\rho =$ const (dotted). The slanted crosses give the distances of the CBF from the eruption center measured toward the solar disk center with the dash-dotted curve showing their fit with $\rho \propto r^{-2}$. (b) Instantaneous velocities computed for a strong shock with $\rho \propto r^{-2}$ (solid) and a weak shock with $\rho = $ const (dotted).}
\label{fig:grechnev2008}
\end{center}
\end{figure*}

\citet{grechnev2008} interpreted CBFs as strong point-like explosions in a variable-density medium \citep[cf.,][]{sedov1981}. They considered propagation of a self-similar blast wave excited by an explosion of an energy $E$ in media (1) with constant density, and (2) with a radial density falloff from the explosion center, $\rho \propto r^{-\alpha}$. For a constant density, $\rho = \rho_0, E = \rho_0 R^3 v^2 =$ constant, where $R$ is the radius and $v$ the velocity of the shock front. Thus, 
\begin{equation}
v = \left(\frac{E}{\rho R^3}\right)^{1/2} \propto R^{-3/2}
\end{equation}
and $R \propto t^{2/5} $. Similarly, for $\rho = br^{-\alpha}$, the velocity can be given as
\begin{equation}
v  \propto R^{[-(3-\alpha)/2]}
\end{equation}
and $R \propto t^{[2/(5-\alpha)]}$. A strong spherical shock decelerates if $\alpha \leq$~3 and accelerates if $\alpha \geq$ ~3. From their analysis, they showed that the resulting kinematics supports their hypotheses that CBFs are coronal blast shocks (see Figure~\ref{fig:grechnev2008} for details).  Their ultimate conclusion, after a comparison with results of Warmuth and co-authors, was that CBFs are probably moderate to strong shocks, which subsequently damp to moderate intensity waves.

\begin{figure*}[!t]
\begin{center}
\includegraphics[width=1\textwidth]{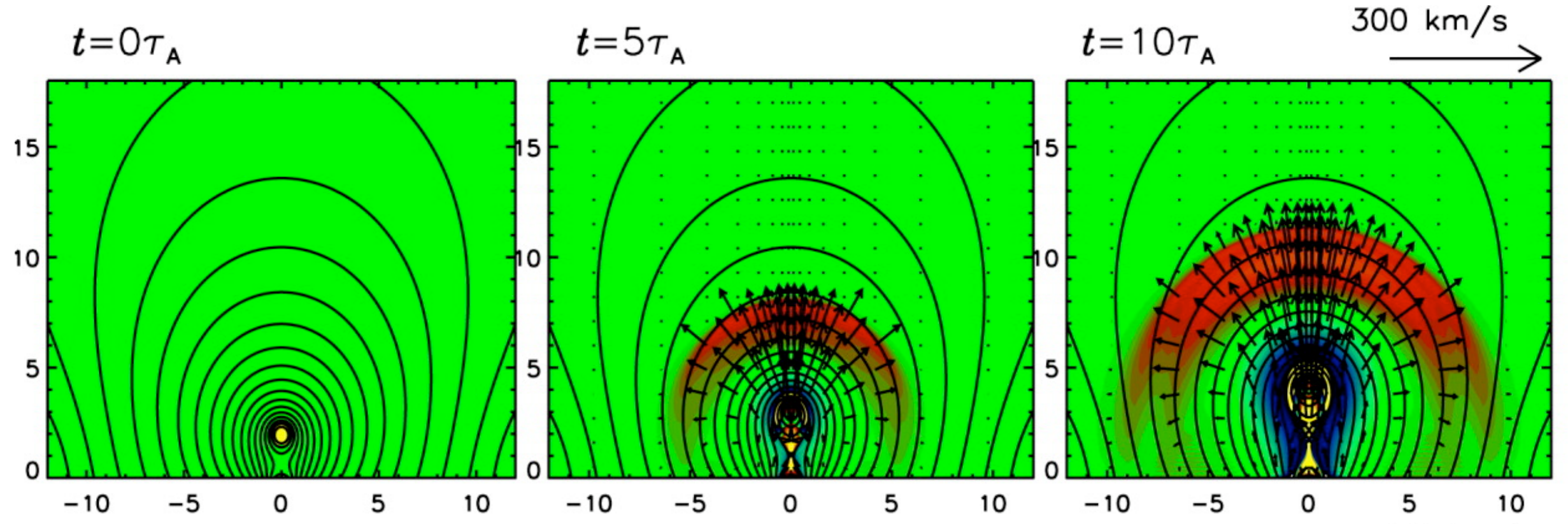}
\includegraphics[width=0.6\textwidth]{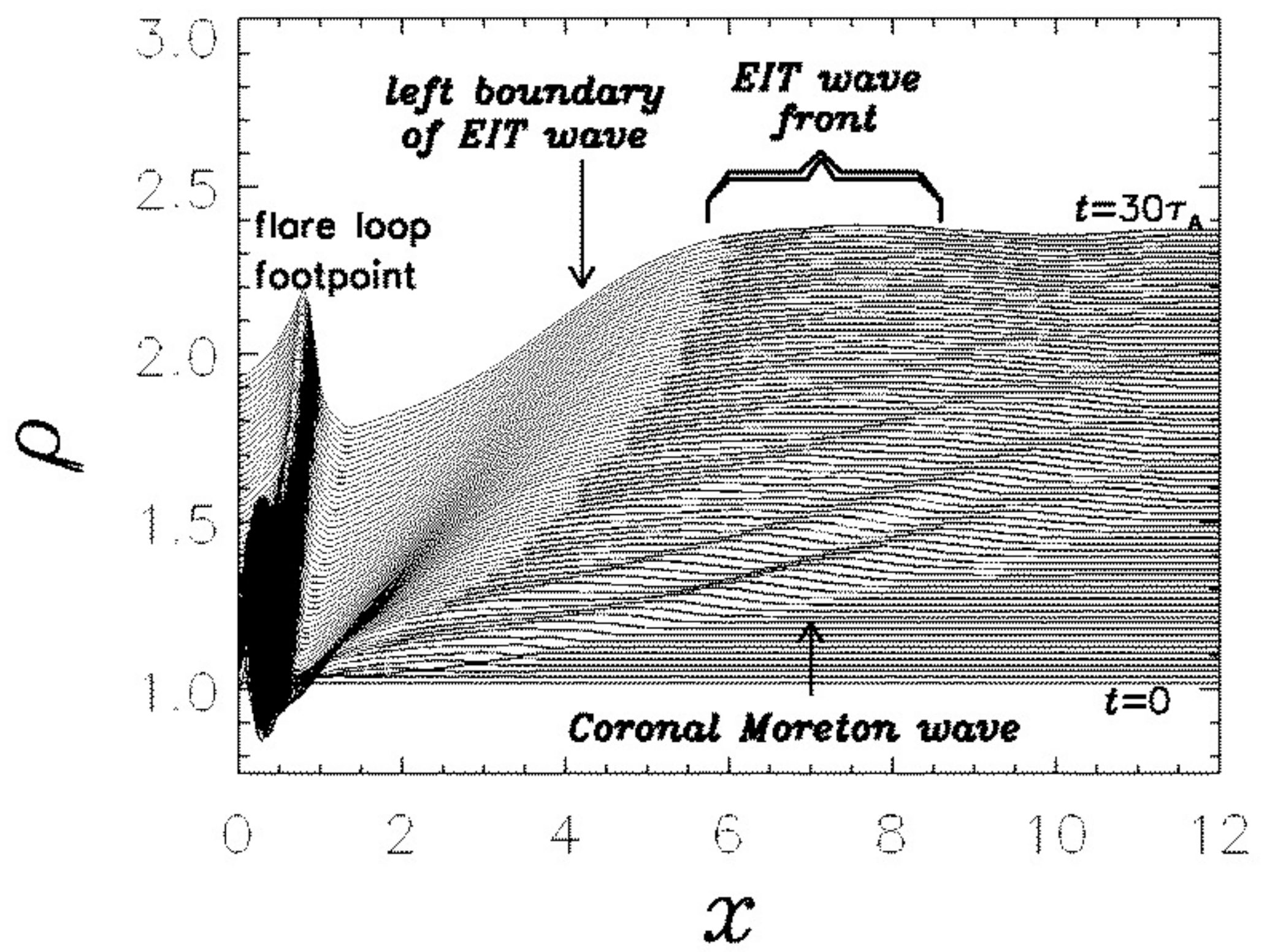}
\caption{{\it Top:} The Chen model for CBFs. As the flux tube rises a shock front is induced in front of it (red in the figure). The legs of this shock front sweep the chromosphere and are observed as Moreton waves. Simultaneously, the CBF is observed as a brightening as the field lines covering the flux rope open. {\it Bottom:} Stack plot showing  the evolution of density ($\rho$) with propagation distance \citep{chen2002}.}
\label{fig:chensimulation}
\end{center}
\end{figure*}

The most recent addition to the wave interpretation of CBFs was put forward by \citet{willsdavey2007}. They examined the nature of CBFs, finding them to be incongruous with fast-mode MHD plane-waves. They suggest that CBFs are in fact more consistent with solitons, that is, waves which maintain constant pulse shape and velocity-dependent amplitudes.  \citet{willsdavey2007} described four major inconsistencies that arise from the fast-mode MHD interpretation of CBFs, namely: velocity magnitudes; plasma~$\beta$; propagation speed variance; and pulse coherence. The two key problems identified by \citet{willsdavey2007} are: 1) wave speeds are too slow for a significant number of observations to be explained using fast-mode MHD waves, and 2) the large range of propagation speed are inconsistent with the wide range of plasma conditions in the solar corona. It should be noted that these issues may be resolved with high cadence ($\leq$1 minute) observations from instruments such as \emph{SDO}/AIA and \emph{Proba-2}/SWAP. \citet{willsdavey2007} suggest that CBF are more consistent with MHD solitons. One key difference between plane wave and soliton solutions is the velocity dependence. With a linear MHD solution, wave speed is determined solely by properties of the transmission medium. Soliton speed is additionally dependent on the amplitude of the pulse. 

An alternative to the purely wave-based interpretation was put forward by \citet{chen2002}. They found evidence for CBFs and Moreton waves in numerical MHD simulations of an erupting flux rope. They showed that as the CME flux rope rises, a piston-driven shock is formed along the envelope of the expanding CME, which sweeps the solar surface as it propagates. They proposed that the legs of the shock produce Moreton waves, while a slower wave-like feature with enhanced leading emission was identified as being an CBF. The results of the simulation may be seen in Figure~\ref{fig:chensimulation}.  They propose that this feature is a CBF resulting from the successive opening of magnetic field lines. Extending on this, \cite{chen2005} and \cite{chenNfang2005} reported on MHD simulations performed to demonstrate how the typical features of CBFs can all be accounted for by successive stretching or opening of closed field lines driven by an erupting flux rope. It was shown that CBFs, which border the expanding dimmming region, were produced by the successive opening (or stretching) of the closed magnetic field lines. H$\alpha$ Moreton waves were found to propagate outward synchronously with the SXR waves, lagging behind the latter spatially. However, the CBFs velocity was found to be approximately a third of the Moreton wave velocity. Furthermore, \cite{chen2005} performed simulations to reproduce the phenomena of stationary CBF fronts, which are located near the footpoint of a magnetic separatrix, consistent with observations \citep{thompson2000}. 

\begin{figure*}[!t]
\begin{center}
\includegraphics[keepaspectratio, width=1\textwidth]{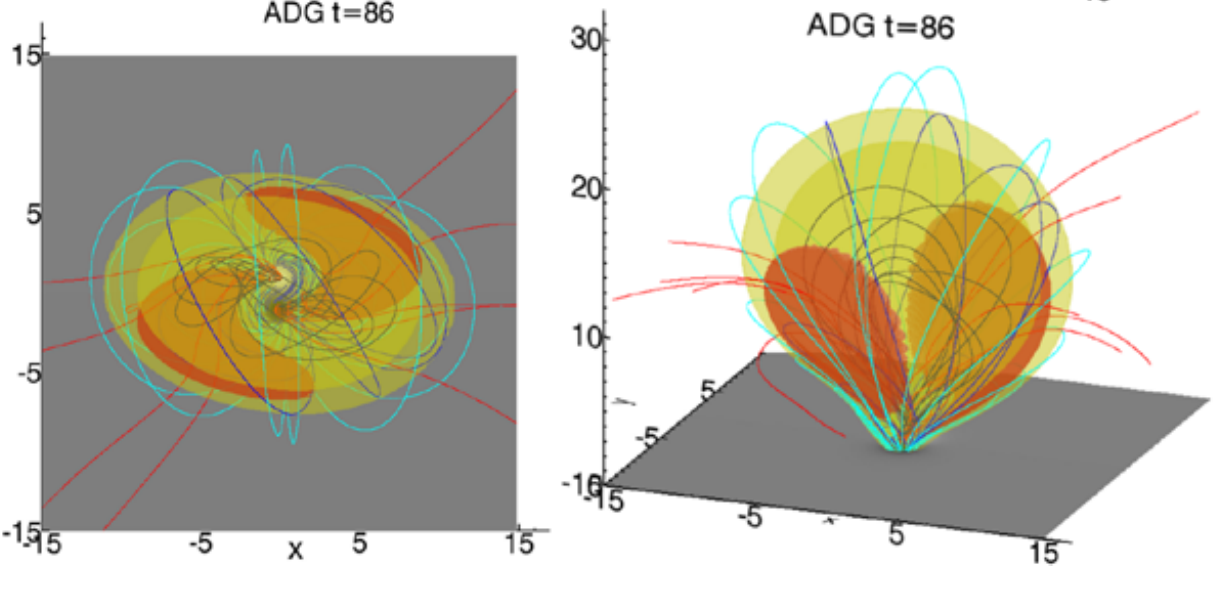}
\caption{The current shell model of \citet{delannee2008}. The current shell is induced at the outer boundary of a rising CME flux tube. Joule heating is visible as a brightening when projected onto the disk.}
\label{fig:delanneemodel}
\end{center}
\end{figure*}

\citet{delannee2000} proposed that large-scale propagating disturbances were more related to changes in magnetic topology of the solar atmosphere due to the eruption of a CME than to a wave. This work was expanded on in \citet{delannee2007} and \citet{delannee2008}. In these papers, CBFs were again explained in terms of changes in magnetic topology as CME legs move through the lower corona. The observed bright fronts of CBFs were proposed to result from Joule heating of coronal plasma at the interface between the CME legs and the surrounding quiet Sun magnetic field. \citet{delannee2008} used two independent 3D MHD codes to perform numerical simulations of a slowly rotating magnetic bipole, which progressively result in the formation of a CME flux-tube ejection. A large-scale and narrow current shell was observed around the twisted flux-tube during the initial stages of its expansion. This current shell was formed by the return currents, which separate the twisted flux tube from the surrounding fields. The current density integrated over the altitude had an elliptical shape with the generic spatial properties of an CBF. The timing, orientation, and location of bright and faint patches observed in the two CBFs were remarkably well reproduced. As a result, they hypothesised that CBFs are the observational signature of Joule heating in electric current shells. An example of the current shell model of CBFs is given in Figure~\ref{fig:delanneemodel}.

\begin{figure*}[!t]
\begin{center}
\includegraphics[width=1\textwidth]{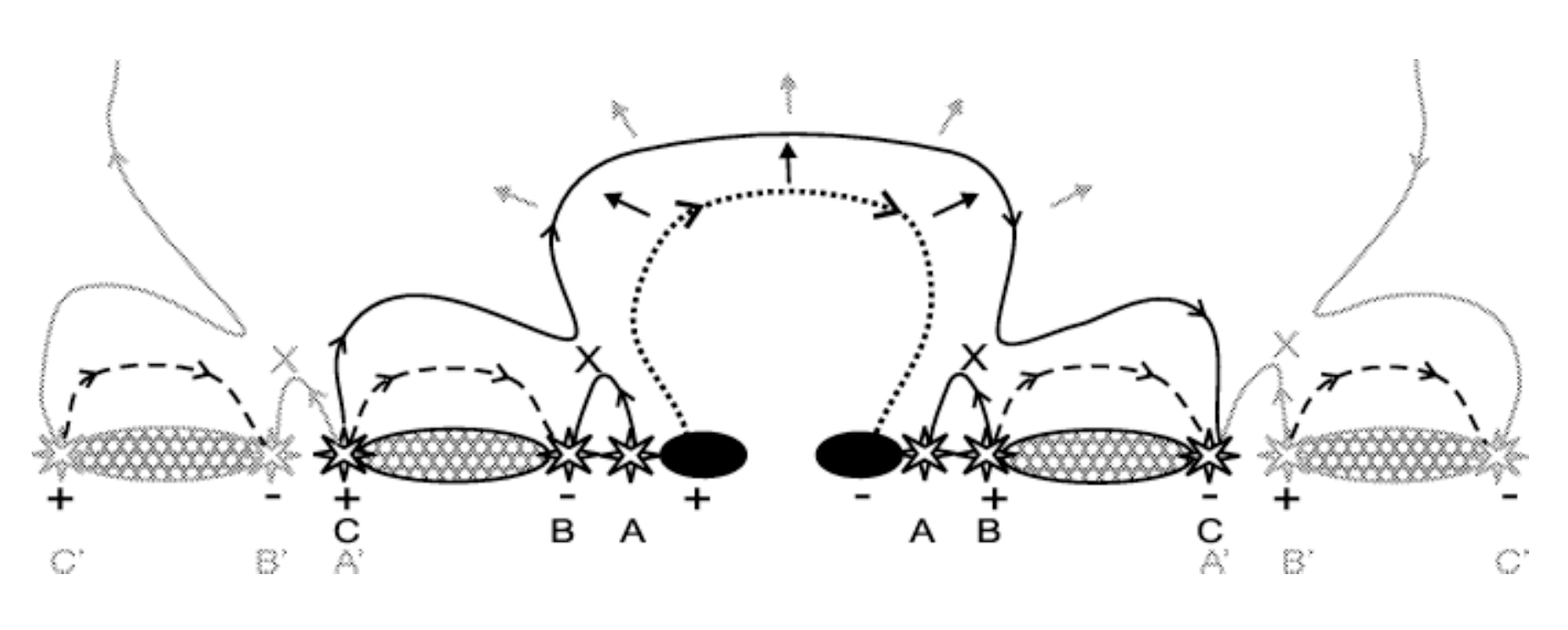}
\caption{The Attrill cartoon for a CBF. As the flux tube rises its magnetic field lines reconnect with quiet Sun loops creating a brightening (as at A and B on the figure). These subsequent reconnections are observed as the propagating wavefront \citep{attrill2007}.}
\label{fig:attrillmodel}
\end{center}
\end{figure*}

The concept of magnetic field re-configuration during CME lift-off and CBF propagation was interpreted in terms of a cartoon proposed by \citet{attrill2006,attrill2007,attrill2009}. In this case, interchange reconnection as opposed to field-line stretching was postulated to account for the properties of CBFs. The Attrill picture suggested that the propagation of the wave front is due to consecutive reconnections in the quiet Sun of favourably orientated magnetic field lines as a magnetic flux tube expands in an active region. With reference to Figure~\ref{fig:attrillmodel}, the expanding CME structure (dotted line) is suggested to reconnect with surrounding favorably orientated quiet Sun loops (dashed lines). The reconnections produce brightenings at points A, B, and C, as a result of gentle chromospheric evaporation. 

The Attrill cartoon is however not without criticism. \citet{delannee2009} examined the validity of this mechanism describing the computed magnetic field topology underlying a coronal wave event studied by \citet{attrill2007}. The active region in question was found to be magnetically linked to regions at a distance of 300~Mm, including the northern coronal hole and the opposite hemisphere, but not to the quiet Sun surrounding the active region \citep{delannee2009}. The outer border of the active region was found to be at the boundary of two different topological magnetic domains, which usually acts as a barrier along which magnetic field lines can slip, but through which they cannot pass. \citet{delannee2009} therefore suggested that the quiet Sun around the active region, in this case, should be barely perturbed by the motion occurring in the active region in such a pre-event magnetic field configuration. These results led \citet{delannee2009} to conclude that, for this event, the quiet Sun should not undergo any reconnection process due to the eruption in the active region studied, in contrast to the Attrill picture.

\begin{figure*}[!t]
\begin{center}
\includegraphics[keepaspectratio, width=0.355\textwidth,clip=,trim=0mm 0mm 40mm 0mm]{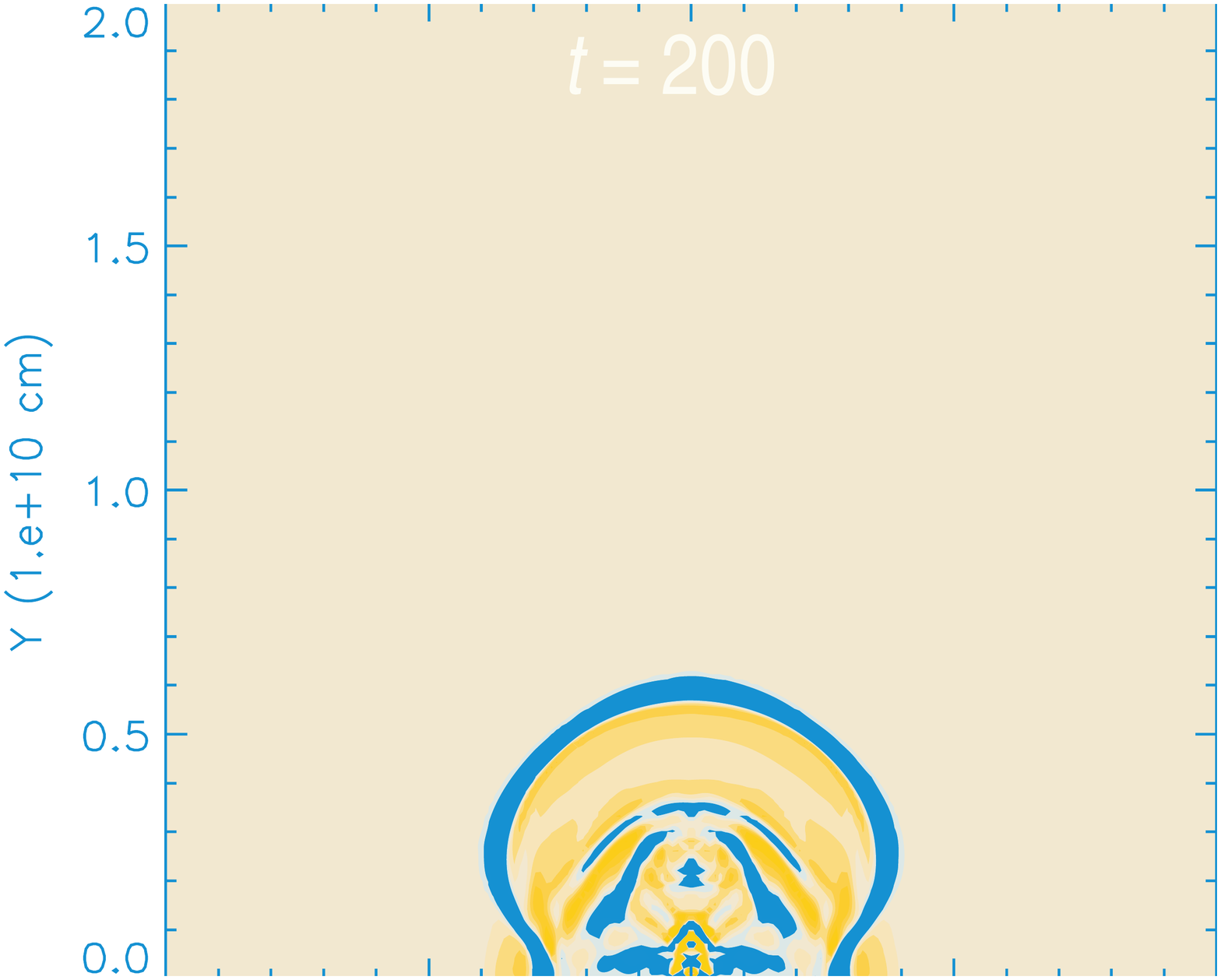}
\includegraphics[keepaspectratio, width=0.3\textwidth,clip=,trim=40mm 0mm 40mm 0mm]{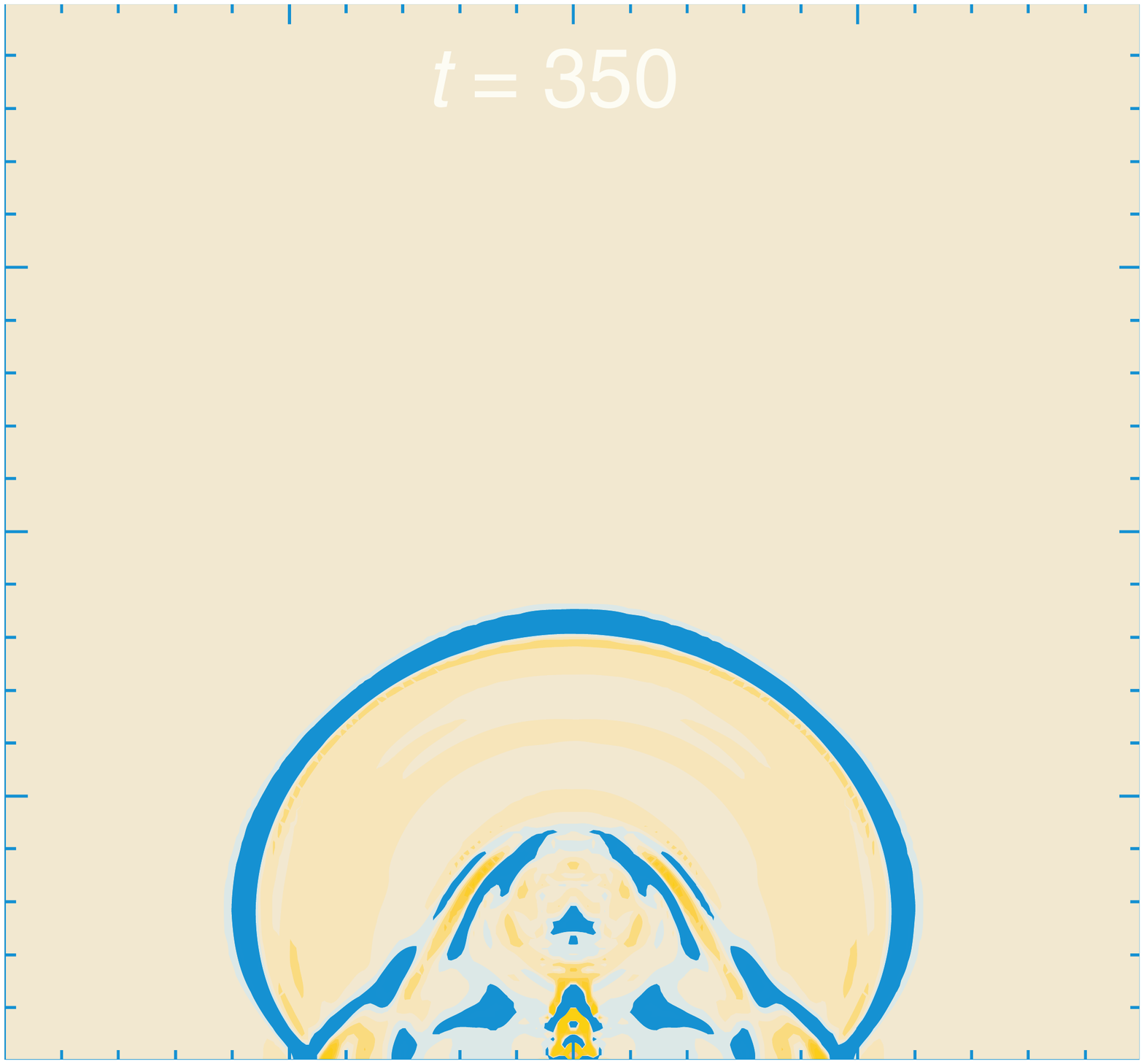}
\includegraphics[keepaspectratio, width=0.3\textwidth,clip=,trim=40mm 0mm 40mm 0mm]{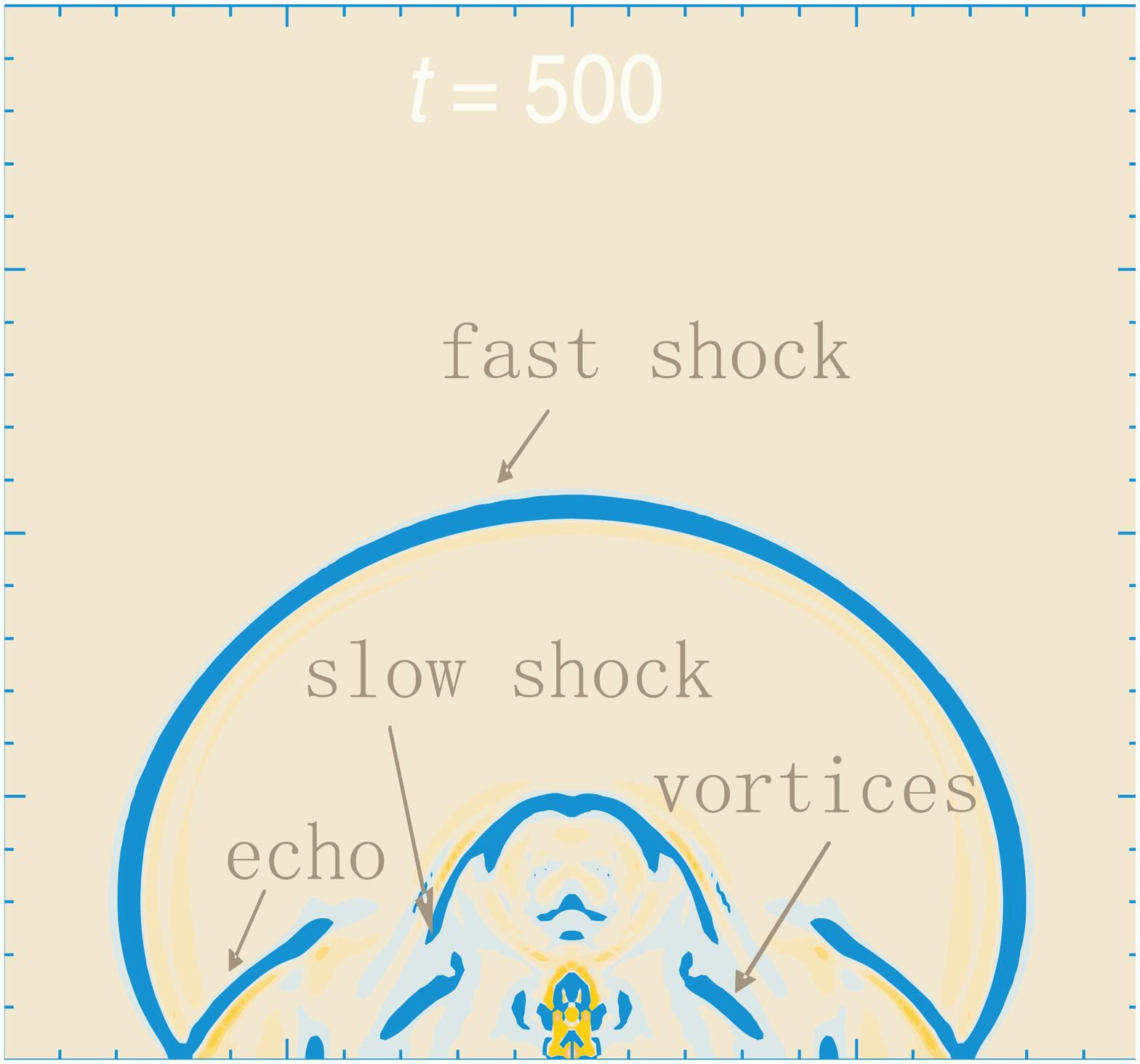}
\caption{Evolution of the velocity divergence in the numerical simulations of \citet{wang2009}. The fast shock, slow shock, echo, and vortices are denoted in the right-most panel, while the time of each images is given in seconds at the top of each panel.}
\label{fig:wangmodel}
\end{center}
\end{figure*}

Most recently,  \citet{cohen2009} and \citet{wang2009} have performed large-scale numerical simulations of flux-rope lift-offs and CBFs. The simulations of \citet{cohen2009} produced a diffuse coronal bright front comparable with observations in the 195~\AA\ passband. They also provides further evidence that CME expansion leads to the opening of coronal field lines on a global scale. Of particular note is that the CME footprint maps directly to the coronal wave in their model. It appears from this and other work \citep[e.g.,][]{zhukov2004} that both wave and non-wave models may be required to explain the observational properties of CBFs. The simulations of \citet{wang2009} introduce an additional degree of complexity to the wave versus non-wave argument. Rapid motions of the erupting flux rope in their simulations produce velocity vortices behind the rope, together with slow- and fast-mode shocks in front of the rope (see Figure~\ref{fig:wangmodel}). The velocity vortices at each side of the flux rope propagate near-horizontally away from their source of excitation with a speed of $\sim$40\% of the associated Moreton wave. These velocities are comparable to those reported by \citet{chen2002} despite the difference in their respective models. 

\section{Conclusions and Future Prospects}
\label{sect:conclusions}

Since the discovery of CBFs in 1997 in \emph{SOHO}/EIT, to the launch of SDO in February 2010, there has been a plethora of papers on the observational characteristics of these phenomena and their theoretical interpretation. These studies have explored a range of additional transient phenomena, such as CMEs, radio bursts, Moreton waves, SEPs, which in some cases have been directly associated with CBFs.  While CBF morphological and kinematical properties are now relatively well described, there are many fundamental questions regarding their physical origin that remain unresolved. These include:
\begin{enumerate}
\item
\emph{How are CBFs launched or excited?} There is recent evidence to suggest that they are excited forward of the expanding flanks of CME \citep{patsourakosNvourlidas2009,veronig2010}, but how can this be accounted for theoretically? Alternatively, are CBFs launched by a piston-like excitation mechanism \citep[e.g.,][]{cliver1999} near the CME launch-site as implied by their close temporal relationship with hard X-ray bursts \citep{veronig2008}?

\item
\emph{What physical mechanism accounts for CBF morphology and kinematics?} Although wave-based models explain many of the observed characteristics, such as pseudo-radial propagation, pulse width, mean intensity, and velocity, they cannot account for stationary CBFs. A reconnection based model might be appropriate appropriate at coronal hole boundaries, but an extremely fast and possibly unphysical reconnection rate would therefore be required to account for the observed quiet-Sun propagation velocities of CBFs.

\item
\emph{How do CBFs related to chomospheric and coronal disturbances?} In particular, are Moreton waves the chromospheric signature of CBFs or are they completely separate phenomena with coincidentally similar characteristics? The former would seems more likely, but is yet to be unambiguously confirmed.

\item
\emph{What is the energy content of CBFs?} The total energy release in a flare/CME is known to be approximately 10$^{25}$~J, with the majority of the energy going into accelerating the CME to velocities of hundreds and sometimes thousands of kilometers per second \citep{emslie2004}. Initial studied have shown that CBFs may have an energy comparable to a nanoflare \citep[$\sim$10$^{18}$~J;][]{ballai2005}, but these results are yet to be corroborated. 
\end{enumerate}

Observationally, accurately determining the velocity of CBFs remains a topic of much discussion, especially considering the implication that an increased average velocity would have for their physical interpretation. The \emph{Proba-2} satellite launched in November 2009 carries onboard an EUV imager, SWAP, which enables images to be obtained with a cadence of 1~minute in the 171~\AA\  passband. This will allow us to gain an insight into the kinematics of CBFs with a sampling rate appropriate to the predicted time-scales of changes in CBF properties. Furthermore, \emph{SDO} carries on-board AIA, which is capable of taking 10 second cadence images in multiple EUV passbands. This will not only enable us to measure CBF velocities with high precision, but to examine the physical properties of the waves at various heights in the solar corona.  

The volume of data anticipated from \emph{Proba-2}/SWAP and \emph{SDO}/AIA in particular highlights the necessity for the development of automated detection methods for various types of solar activity, including CBFs. To date, CBFs have been detected by eye using a point-and-click methodology. This will not be possible with AIA, though, and so automated image processing techniques will need to be developed to detect and track CBFs as they propagate through various layers in the solar atmosphere. Examples of automated techniques to detect EUV dimmings and bright fronts include \citet{podladchikova2005,willsdavey2006,attrillNwillsdavey2009}. The NEMO system\footnote{http://sidc.oma.be/nemo/}, developed by \citet{podladchikova2005}, autonomously detects solar eruptions in image sequences from EIT. NEMO consists of a series of high level image processing techniques developed to extract eruptive features from the EUV solar disk under complex solar conditions and is based on the general statistical properties and the underlying physics of eruptive on-disk events. One of the specific NEMO features is the capability to find a CBF in the large family of solar dimmings. \citet{attrillNwillsdavey2009} extended the ideas of NEMO to  develop automated coronal-dimming region detection and extraction algorithm to determine physical properties of CBFs, such as spatial location, area, and volume. These methods will be used to mine \emph{SDO}/AIA images in near-realtime and not only detect the occurrence of CBFs but also facilitate the automated characterisation of their kinematic and morphological properties.

The physical interpretation of CBFs remains controversial, with many in the community being split between the wave and pseudo-wave theories. Each of these have their merits, but neither can completely account for the detailed morphological and kinematic properties of CBFs and their relationship with other solar phenomena, such as flares and CMEs. As noted by \citet{thompson2000}, no one theory can as yet accurately describe all observed properties of EIT and Moreton waves. This could result in the actual solution being an amalgamation of several distinct theories. 

\begin{acknowledgements}
DML is a Government of Ireland Scholar, supported by the Irish Research Council for Science, Engineering and Technology (IRCSET).
\end{acknowledgements}

\end{document}